\documentclass[smallextended]{svjour3}       
\smartqed  

\usepackage[utf8]{inputenc} 
\usepackage[T1]{fontenc}
\usepackage{graphicx}
\usepackage{color} 
\usepackage[bookmarksdepth=2]{hyperref} 
\usepackage[misc,geometry]{ifsym}  
\usepackage{float} 
\usepackage[sort&compress,square,comma,numbers]{natbib}
\usepackage{amsmath} 
\usepackage[table,xcdraw]{xcolor} 
\usepackage{mdframed} 
\usepackage{xspace} 
\usepackage{color,soul} 
\usepackage{framed}
\usepackage{subcaption}
\usepackage{enumitem}
\usepackage[many]{tcolorbox}
\usepackage{rotating}
\usepackage{tabularx}
\usepackage{spverbatim}
\usepackage{siunitx}
\usepackage{booktabs}
\usepackage{listings}
\usepackage{textcomp} 
\usepackage{pbox}
\usepackage{url}
\usepackage{moresize}

\usepackage{multirow} 
\usepackage[normalem]{ulem}
\useunder{\uline}{\ul}{}
\usepackage{longtable}

\newcommand{\dapp}{\DH App\xspace}
\newcommand{\dapps}{\DH Apps\xspace}

\makeatletter

\renewenvironment{quote}
  {\list{}{\rightmargin=0.0cm \leftmargin=0.52cm}%
   \item\relax}
  {\endlist}
 
\patchcmd{\NAT@citex}
{\@citea\NAT@hyper@{%
		\NAT@nmfmt{\NAT@nm}%
		\hyper@natlinkbreak{\NAT@aysep\NAT@spacechar}{\@citeb\@extra@b@citeb}%
		\NAT@date}}
{\@citea\NAT@nmfmt{\NAT@nm}%
	\NAT@aysep\NAT@spacechar\NAT@hyper@{\NAT@date}}{}{}

\patchcmd{\NAT@citex}
{\@citea\NAT@hyper@{%
		\NAT@nmfmt{\NAT@nm}%
		\hyper@natlinkbreak{\NAT@spacechar\NAT@@open\if*#1*\else#1\NAT@spacechar\fi}%
		{\@citeb\@extra@b@citeb}%
		\NAT@date}}
{\@citea\NAT@nmfmt{\NAT@nm}%
	\NAT@spacechar\NAT@@open\if*#1*\else#1\NAT@spacechar\fi\NAT@hyper@{\NAT@date}}
{}{}
\makeatother

\renewcommand\paragraph{\@startsection{paragraph}{4}{\z@}%
	{-2.5ex\@plus -1ex \@minus -.25ex}%
	{1.25ex \@plus .25ex}%
	{\normalfont\normalsize\itshape}}
\let\oldunderscore\_
\renewcommand\_{\ifmmode \oldunderscore \else \textunderscore\allowbreak \fi}

\newtcolorbox{mybox}[2][]{
top=0.15in,left=4pt,right=4pt,bottom=4pt,
fonttitle=\bfseries,
colbacktitle=gray,
colback=gray!5,
colframe=gray!40!black,
enhanced,
attach boxed title to top left={xshift=1.5em,yshift=-\tcboxedtitleheight/2},
boxed title style={size=small},
drop shadow={black!50!white},
title=#2,#1}

\newcommand{\countobservations}{
    \def \countobservations{1}
}
\newcounter{observation}
\newenvironment{observation}[1]
{   \refstepcounter{observation}
    \noindent \textbf{Observation~\theobservation) \textit{#1}}
}

\countobservations

\newcommand{\countimplications}{
    \def \countimplications{1}
}
\newcounter{implication}
\newenvironment{implication}[1]
{
    \refstepcounter{implication}
    \noindent \textbf{Implication~\theimplication ) \textit{#1}}
}

\countimplications

\newcommand\RQOne{How well do blockchain internal factors characterize transaction processing times?}
\newcommand\RQTwo{How well does gas pricing behavior characterize transaction processing times?}

\def\bitcoin{%
	\leavevmode
	\vtop{\offinterlineskip 
		\setbox0=\hbox{B}%
		\setbox2=\hbox to\wd0{\hfil\hskip-.03em
			\vrule height .3ex width .15ex\hskip .08em
			\vrule height .3ex width .15ex\hfil}
		\vbox{\copy2\box0}\box2}}

\graphicspath{{images/}}

\begin{document}
	
	\title{What makes Ethereum blockchain transactions be processed fast or slow? An empirical study}

	\author{Michael~Pacheco \and Gustavo~A.~Oliva \and Gopi~Krishnan~Rajbahadur \and Ahmed~E.~Hassan}

	\institute{
		\Letter \space Michael Pacheco, Gustavo A. Oliva, and Ahmed E. Hassan \at
		Software Analysis and Intelligence Lab (SAIL), School of Computing \\
		Queen's University, Kingston, Ontario, Canada\\    
		\email{\{gustavo,ahmed\}@cs.queensu.ca} \and		
		Gopi Krishnan Rajbahadur \at
		Centre for Software Excellence (CSE), Huawei, Canada \\
		\email{gopi.krishnan.rajbahadur1@huawei.com}
	}


	\date{Received: date / Accepted: date}

	\maketitle

	\begin{abstract}
The Ethereum platform allows developers to implement and deploy applications called \dapps onto the blockchain for public use through the use of smart contracts. To execute code within a smart contract, a paid transaction must be issued towards one of the functions that are exposed in the interface of a contract. However, such a transaction is only processed once one of the miners in the peer-to-peer network selects it, adds it to a block, and appends that block to the blockchain This creates a delay between transaction submission and code execution. It is crucial for \dapp developers to be able to precisely estimate when transactions will be processed, since this allows them to define and provide a certain Quality of Service (QoS) level (e.g., 95\% of the transactions processed within 1 minute). However, the impact that different factors have on these times have not yet been studied. Processing time estimation services are used by \dapp developers to achieve predefined QoS. Yet, these services offer minimal insights into what factors impact processing times. Considering the vast amount of data that surrounds the Ethereum blockchain, changes in processing times are hard for \dapp developers to predict, making it difficult to maintain said QoS. In our study, we build random forest models to understand the factors that are associated with transaction processing times. We engineer several features that capture blockchain internal factors, as well as gas pricing behaviors of transaction issuers. By interpreting our models, we conclude that features surrounding gas pricing behaviors are very strongly associated with transaction processing times. Based on our empirical results, we provide \dapp developers with concrete insights that can help them provide and maintain high levels of QoS.

	\end{abstract}

	\section{Introduction}
\label{sec:intro}

A blockchain provides a secure and decentralized infrastructure for the execution and record-keeping of digital transactions. A \textit{programmable} blockchain platform supports \textit{smart contracts}, which are stateful, general-purpose computer programs that enable the development of \textit{blockchain-powered applications}. Ethereum is currently the most popular programmable blockchain platform. In Ethereum, these blockchain-powered applications are known as decentralized applications (\dapps).

Transactions are the means through which one interacts with Ethereum. Ethereum defines two types of transactions: \textit{regular} transactions and \textit{contract} transactions. Regular transactions enable transfers of cryptocurrency (Ether, or simply ETH) between end-user accounts. In turn, contract transactions invoke functions defined in the interface of a smart contract. When an end-user triggers a certain functionality on the frontend of a \dapp (e.g., checkout the items in the shopping cart of an online shopping application), such a request is translated into one or more blockchain contract transactions. \dapps themselves facilitate these transactions by submitting them to the blockchain (e.g., Ethereum). That is, end-users are completely unaware that a blockchain is in use, as transactions are handled entirely by the \dapp.

In Ethereum, all transactions must be paid for, including contract transactions. However, transaction prices are not prespecified. Transaction issuers need to decide how much they wish to pay for each transaction by the assignment of the \textit{gas price} parameter, paid in Ether (ETH). The higher one pays for a transaction, the faster it is generally processed. This is because there is a financial incentive that \textit{mining nodes} (i.e., those who choose, prioritize, and process all transactions on Ethereum) receive for transactions that they process and verify. This incentive is a function of the gas prices that are associated with the transactions they choose to process. As a result, the execution of code on programmable blockchains such as Ethereum takes time, as the corresponding transaction(s) must first be chosen by the mining node that will append the next block to the blockchain. Consistently setting up transactions with high gas prices is not an option, since it would render the \dapp economically unviable. Hence, for a \dapp to be profitable, it needs to issue transactions with the lowest gas price that fulfills a predefined Quality of Service (QoS) (e.g., transactions processed within 1 minute in 95\% of the cases). To achieve such a goal, \dapp developers make use of \textit{processing time estimation services}. 

Although existing processing time estimation services are useful, they offer little to no insight into what drives transaction processing times in Ethereum (except for gas prices). This happens either because the underlying machine learning model that is used by these services is a complete black-box (e.g., Etherscan's Gas Tracker\footnote{\url{https://etherscan.io/gastracker}}) or because the service does not clearly indicate the features that drive the model's decisions (e.g., EthGasStation\footnote{\url{https://ethgasstation.info}}). The processing time estimations made by these services impact the gas prices of transactions sent by \dapp developers. We thus argue that these services should be interpretable. Without explanations or insights into the underlying models, the estimations provided by these services lack reliability, robustness, and trustworthiness.

The very limited insights brought by current estimation services make \textit{changes} in processing times virtually unpredictable, since \dapp developers do not know exactly what metric(s) to track. As a consequence, \dapp developers struggle to define proper gas prices and manage transaction submission workloads. For instance, if overall gas prices are high, but those are likely to lower soon, then \dapp developers might be better off holding their transactions for a while and submitting them later on in batches. Since no processing time estimation service offers such information, \dapp developers are forced to blindly rely on these services, and monitor and analyze information on the Ethereum blockchain to discover impacting factors themselves in order to predict transaction processing times in the future. This issue calls for a unified, free, and open-source alternative to the existing services, which would provide DApp developers with explainable processing time models and predictions.

Building an explainable transaction processing time model entails discovering and understanding the factors that influence (or are at least correlated with) processing times. The resulting information surrounding the model will allow \dapp developers to make more informed decisions about what metrics they choose to monitor in real-time in order to set adequate gas prices. As a first attempt at bridging this gap, in this study we set out to determine and analyze the factors that are associated with transaction processing times in Ethereum. In the following, we list our research questions and the key results that we obtained:

\smallskip \noindent \textbf{RQ1: \RQOne} To answer this RQ, we first build a random forest model with an extensive set of features which capture internal factors of the blockchain. (e.g., the difficulty to mine a block, or the characteristics of the contracts and transactions) of the Ethereum blockchain. These features are derived directly from the contextual, behavioral, and historical characteristics associated with 1.8M transactions. Next, we interpret the model using adjusted $R^2$ and SHAP values. 

\smallskip \noindent \textit{Our model achieves an adjusted $R^2$ of 0.16, indicating that blockchain internal factors explain little of the variance in processing times.}

\smallskip \noindent \textbf{RQ2: \RQTwo}
Similar to the stock market, the Ethereum blockchain ecosystem is impacted by a plethora of external factors, including speculation, the global geopolitical situation, and cryptocurrency regulations. Many of these factors are largely unpredictable and difficult to be captured in the form of engineered features. Yet, all these factors are known to impact the \textit{gas pricing behavior} of a large portion of transaction issuers. Hence, in RQ2 we build a model that aims to account for the gas pricing behavior of transaction issuers (i.e., the mass behavior). We do so by engineering additional features that indicate how the price of a given transaction compare to that of recently submitted transactions.

\smallskip \noindent \textit{Our model achieves an adjusted $R^2$ score of 0.53 (2.3x higher than the baseline model), indicating that it explains a substantially larger amount of the variance in processing times. The most important feature of our model is the median percentage of transactions processed per block, in the previous 120 blocks (\textasciitilde30min ago), with a gas price below that of the current transaction.}


\smallskip \noindent \textbf{Paper organization.} The remainder of this paper is organized as follows. Section~\ref{sec:methodology} describes our study methodology, including the data collection, our engineered feature-set, and the model construction. Sections~\ref{sec:rq1} and~\ref{sec:rq2} show the motivation, approach, and findings associated with the research questions addressed in this paper. Section~\ref{sec:related-work} discusses related work. Section \ref{sec:threats} discusses the threats to the validity of our study. Finally, Section \ref{sec:conclusion} states our concluding remarks. Appendix~\ref{appendix:background} provides a summary of the key blockchain concepts that we employ throughout this paper.
	\section{Methodology}
\label{sec:methodology}

We build a machine learning model with a comprehensive feature-set in order to determine the factors that are strongly associated with the processing times of transactions in Ethereum. In this section, we explain our approaches to data collection (Section~\ref{sec:data-collection}), feature engineering (Section~\ref{sec:feature-set}), and model construction (Section~\ref{sec:model-construction}). 

\subsection{Data Collection}
\label{sec:data-collection}

In this section, we describe the data sources that we used (Section~\ref{sec:data-sources}) and the approach that we followed to collect the data (Section~\ref{sec:data-steps}).

\subsubsection{Data Sources}
\label{sec:data-sources}

Our study relies on two data sources, namely Etherscan and Google BigQuery:

\smallskip \noindent \textbf{Etherscan.} Etherscan\footnote{https://etherscan.io/} is a popular real-time dashboard for the Ethereum blockchain platform. Etherscan has its own nodes in the blockchain, which are used to actively monitor the activity of the network in real time. This includes monitoring various details regarding transactions, blocks, and smart contracts. We use the data from this website to obtain transaction processing times (our dependent variable) and to engineer several features used in our model (e.g., pending pool size and network utilization level). Data from Etherscan has been used in several blockchain empirical studies (\cite{Oliva}, \cite{Oliveira21}, \cite{Pierro})).

\smallskip \noindent \textbf{Google BigQuery.} Google BigQuery\footnote{https://cloud.google.com/bigquery} is an online data warehouse that supports querying through SQL. In particular, Google also actively maintains several public datasets, including an Ethereum dataset\footnote{The dataset is \texttt{bigquery-public-data.crypto\_ethereum}}. This dataset is updated daily and contains metadata for several blockchain elements, including transactions, blocks, and smart contracts. Most of the features that we engineer rely on data collected from this dataset.

\subsubsection{Approach} 
\label{sec:data-steps}

\begin{figure}[!htbp]
\centering
    \includegraphics[width=1.0\linewidth]{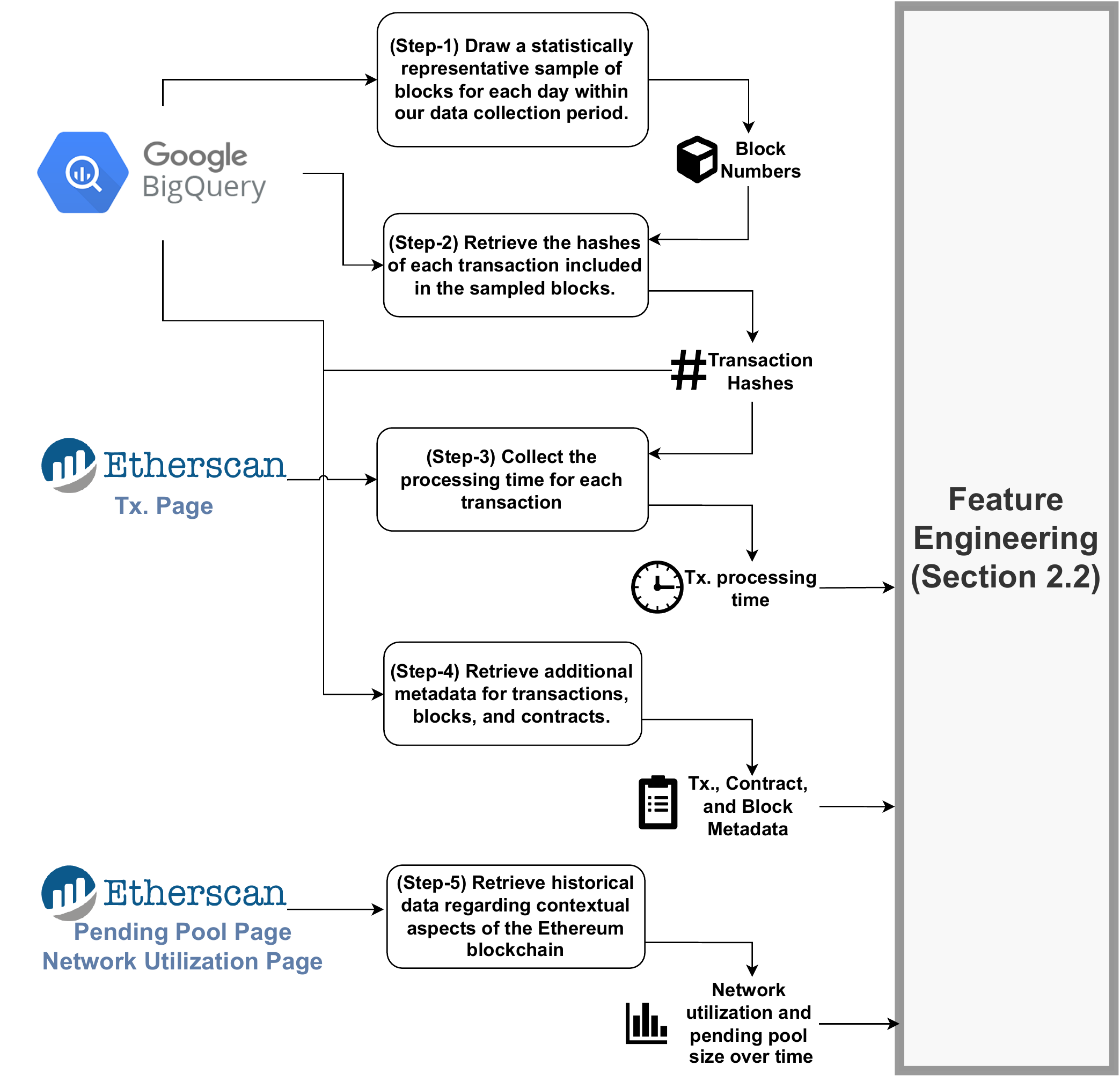}
    \caption{An overview of our data collection approach. The dashed lines represent a connection to a data source.}
    \label{fig:data-collection}
\end{figure}

Figure \ref{fig:data-collection} provides an overview of our data collection approach. In the following, we discuss each of the data collection steps.

\smallskip \noindent \textbf{(Step-1) Draw a statistically representative sample of blocks for each day within our data collection period.} 
To bring data size to more manageable levels, we draw a statistically representative sample of mined blocks for each day within a one-month period. We use a 95\% confidence level and $\pm$ 5 confidence interval, similarly to several prior studies \citep{Ankur2022,Zhang2019,Boslaugh2008}. Following this sampling approach, we obtain a total of 10,865 blocks (362 blocks per day on average). A replication package containing our studied data is available online\footnote{\url{https://drive.google.com/drive/folders/1nyPY3w8TOjne8RItv2TqfvJrfnnfbtI4?usp=sharing}}

\smallskip \noindent \textbf{(Step-2) Retrieve the hashes of each transaction included in the sampled blocks.} We proceed by collecting and saving the \textit{transaction hashes} of all transactions that have been included within the sampled blocks. The \textit{transaction hash} is a unique identifier used to differentiate between transactions. We use these transaction identifiers in several of the data collection steps to collect various associated metadata with each of these transactions. We collect the \textit{transaction hash} for all 1,825,042 (1.8M) transactions included in the 10,865 sampled blocks based on the previous step.

\smallskip \noindent \textbf{(Step-3) Collect the \textit{processing time} for each transaction.} Next we begin to collect the \textit{processing time} provided by the Etherscan platform. For a given transaction \textit{t}, the \textit{processing time} is the estimated difference between the time \textit{t} has been included in a block that has been appended to the blockchain, and the time the user executed the submission of \textit{t} onto the blockchain. We retrieve this data by using the \textit{transaction hash} of each transaction in our set to access the corresponding \textit{processing time} details provided by Etherscan (defined as \textit{confirmation time} by Etherscan). 

\smallskip \noindent \textbf{(Step-4) Retrieve additional metadata for transactions, blocks, and contracts.} We then leverage Google BigQuery to retrieve additional information regarding transactions, including information about the blocks and smart contracts they are associated with (when applicable). Examples of the types of the collected metadata from this source includes details such as gas price, block number, and network congestion. These metadata are obtained by querying the \textit{transactions}, \textit{blocks}, and \textit{contracts} tables.

\smallskip \noindent \textbf{(Step-5) Retrieve historical data regarding contextual aspects of the Ethereum blockchain.} In addition to the \textit{processing time}, the Etherscan platform provides historical data that describes some of the contextual information of the Ethereum blockchain. In particular, we collect data which describes: 1) the amount of transactions in the pending pool\footnote{\url{https://etherscan.io/chart/pendingtx}}, and 2) the network utilization\footnote{\url{https://etherscan.io/chart/networkutilization}}. We then map this set of data to each transaction, associating the relevant context of the blockchain to the time each transaction was executed.
\subsection{Feature Engineering}
\label{sec:feature-set}

To understand the features that impact the transaction processing times, we collect and engineer a variety of features. We explain the rationale behind focusing on these dimensions, as well as the rationale for designing each of the features in our set. Features were carefully crafted based on our scientific ~\citep{Oliva,Kondo2020,Oliva21,Zarir21} and practical knowledge of Ethereum (e.g., buying/selling/trading ETH and watching the cryptocurrency market fluctuations). We list each of the features that we use for building our models in Table~\ref{tab:features} and Table~\ref{tab:add-features}. The definitions of these features touch on several key blockchain concepts. To assist in the comprehension of the description and rationale of our features, we refer the reader to Appendix~\ref{appendix:background} (background on blockchain key concepts).

\smallskip \noindent \textbf{1. Features that capture internal factors} We first engineer features that aim to capture internal factors of the Ethereum blockchain (RQ1), which are summarized in Table~\ref{tab:features}. These features span across a few dimensions, including contextual, behavioral, and historical dimensions. We explain these dimensions below.

\smallskip \noindent \textit{-- Contextual.} Contextual data describes the circumstances in which a given transaction is being processed. This information can be associated with the entire blockchain network, and also the transactions themselves. For example, the network utilization at the time a transaction was executed is a contextual feature. The network utilization should help provide indicators of changes to transaction processing times, such as increase in network traffic causing processing times to be delayed. Similarly, a feature which defines the smart contract a transaction had interacted with in its lifetime also provides contextual insights.

\smallskip \noindent \textit{-- Behavioral.} Behavioral features include information which relate to the user and the particular values that were chosen for each of the transaction parameters. Depending on the experience of the user who is sending the transaction, the values set for these parameters might impact transaction processing times. Most notably, miners have the freedom to prioritize transactions to process using their own algorithms. However, it is common for miners to prioritize processing transactions with higher gas price values, as they receive a reward for processing a particular transaction based on this value. An inexperienced user who is not familiar with gas price values might set too low of a value, resulting in an extremely long processing time. 

\smallskip \noindent \textit{-- Historical.} Historical features relate to features based on historical data from transactions that have already been processed in the blockchain. Such metrics are widely used in machine learning for predictive modelling tasks, as data of the past is used to predict outcomes of the future. For example, a feature which describes the average processing time for processed transactions in the previous $x$ blocks can be used by models to understand and predict processing times for similar transactions in the future. 

\smallskip \noindent \textbf{2. Features that capture changes in gas pricing behavior.} We also engineer features that aim to capture pricing behavior and changes in thereof (RQ2). Our features primarily involve comparing the gas prices of transactions processed in recent blocks with that of the current transaction at hand. These features and their associated rationale are summarized in Table~\ref{tab:add-features}.

\begin{table*}
	\centering
	\caption[The list of features used to capture internal factors of the Ethereum blockchain.]{The list of features used to capture internal factors of the Ethereum blockchain, along with their rationale. These features are used to build our models in~\ref{sec:rq1} in order to predict transaction (tx) processing times. The $\bullet$ symbol represents data retrieved from Etherscan, while the $\star$ symbol represents data retrieved from Google BigQuery. Braces (\{\}) group related features.}
	\label{tab:features}
	\scriptsize
	\renewcommand{\arraystretch}{1.4}
	\resizebox{1.35\textwidth}{!}{%
	\begin{tabular}{lp{4.0cm}p{5.8cm}p{5.8cm}}
		\midrule
		\textbf{Type} &
		\textbf{Name} &
		\textbf{Description} &
		\textbf{Rationale} \\ \midrule
		
		Contextual &
		\texttt{contract\_block\_number}$\star$ &
		The block number of the smart contract that the tx has interacted with. &
		Contracts that have been deployed onto the Ethereum blockchain long ago might impact processing times by containing inefficient code, security vulnerabilities, or bugs, causing them to be avoided by miners. \\ \cmidrule{2-4} 
		&
		\texttt{contract\_bytecode\_length}$\star$ &
		The bytecode length of the smart contract the tx has interacted with. &
		Larger smart contract sizes might result in an increase in execution time, which might also impact processing times. \\ \cmidrule{2-4} 
		&
		\texttt{is\_\{erc721, erc20\}}$\star$ &
		A flag indicating whether the smart contract the tx has interacted with is an ERC721 or ERC20 (token-related) contract. &
		\multirow{2}{=}{Contract txs. might be prioritized by miners, and might also have longer execution times, compared to user txs.} \\
		&
		\texttt{to\_contract}$\star$ &
		A flag indicating whether the tx is interacting with a contract. &
		\\ \cmidrule{2-4} 
		&
		\texttt{pending\_pool}$\bullet$ &
		The total amount of txs. in the pending pool that have yet to be processed, at the time closest to the time of sending the tx. &
		\multirow{2}{=}{The amount of txs. in the pending pool, along with network utilization, reflects the network traffic at a given point in time. We predict an increase in these metrics should also result in increased tx processing times.} \\
		&
		\texttt{net\_util}$\bullet$ &
		The network utilization of the network, as calculated by Etherscan. &
		\\ \cmidrule{2-4} 
		&
		\texttt{day}$\star$ &
		The day of the week the tx was executed on. &
		\multirow{2}{=}{Contextual patterns of the network, such as increased usage, might revolve around certain days of the week or hours in the day.} \\
		&
		\texttt{hour}$\star$ &
		The hour of the day the tx was executed on. &
		\\ \midrule
		Behavioral &
		\texttt{gas\_price\_gwei}$\star$ &
		The gas price value in GWEI set by the user of the tx. &
		The gas price is directly related to tx processing times - miners receive a monetary reward based on the gas prices of the txs. inside the blocks that they append to the chain. As a result, txs. with higher gas prices should have fast processing times. \\
		&
		\texttt{tx\_nonce}$\star$ &
		The total number of txs. sent by the user up until the current tx. &
		User inexperience might be associated with setting inadequate tx parameters, most importantly the gas price. Also, txs. with different nonce values might also be prioritized differently by various miners. \\
		&
		\texttt{value}$\star$ &
		The value of ETH the user is sending along with their tx. &
		Differences in the ETH value might be prioritized differently by various miners. \\
		&
		\texttt{tx\_gas\_limit}$\star$ &
		The gas limit value set by the user of the tx. &
		Difference in gas limits might be prioritized differently by various miners. For example, miners might avoid specific txs. if they think it might result in out of gas error. \\
		&
		\texttt{input\_length}$\star$ &
		The length of the input data associated with the tx. &
		Longer inputs might take longer to process in smart contracts, leading to longer execution time, and also longer processing time. \\ \midrule
		
		Historical &
		\texttt{\{total, avg, med, std\}\_txs\_120}$\star$ &
		The total, average, median, and standard deviation of all processed txs. across all previous 120 blocks, and the previous block. &
		Increases in the number of txs. being processed in the recent past is likely to increase processing times in the near future. \\ \cmidrule{2-4} 
		&
		\texttt{total\_txs\_1}$\star$ &
		The total number of txs. processed in the previous block. &
		\\ \cmidrule{2-4} 
		&
		\texttt{avg\_gas\_price\_gwei\_prev\_day}$\bullet$ &
		The average gas price of txs. processed the previous day. &
		\\ \cmidrule{2-4} 
		&
		\texttt{\{avg, med, std\}\_difficulty\_120}$\star$ &
		The average, median, and standard deviation of the difficulty of the previous 120 blocks. &
		\multirow{2}{=}{The difficulty is directly related to the time taken to process blocks. This value is altered based on the time taken by miners to append blocks to the Ethereum blockchain.} \\
		&
		\texttt{difficulty\_1}$\star$ &
		The difficulty of the previous block. &
		\\ \cmidrule{2-4} 
		&
		\texttt{\{avg, med, std\}\_func\_gas\_usage\_120}$\star$ &
		The average, median, and standard deviation of gas usage of the processed txs. in the previous 120 blocks, targeting the same smart contract function as the current tx. &
		The gas usage from the smart contract txs. are constantly targeting may indicate the popularity and utilization of such contract, which may be prioritized by miners in a particular way. \\ \cmidrule{2-4} 
		&
		\texttt{\{avg, med, std\}\_pending\_pool\_120}$\bullet$ &
		The average, median, and standard deviation of the number of txs. in the pending pool per minute, during the appending of the previous 120 blocks. &
		As the amount of txs. in the pending pool increases, miners have more txs. to prioritize, which should increase processing times. \\ \cmidrule{2-4}
		&
		\texttt{\{avg, med, std\}\_pend\_prices} $\star$ &
		The average, median, standard deviation of gas prices (GWEI) of txs from the same user that are currently in the pending pool and not processed at the same time of the recorded tx For the first tx, the value of these features is zero.
		& 
		The more txs a user has that are simultaneously waiting to be processed, the longer it could take for a single tx of that user to be processed due to the gas prices and nonce of the txs. 
		\\ 
		&
		\texttt{num\_pending}$\star$ &
		The number of txs from the same user that are currently in the pending pool and not processed at the same time of the recorded tx.
		&
		\\ \midrule
	\end{tabular} 
	}%
\end{table*}

\begin{table*}
	\centering
	\caption[The list of features used to capture gas pricing behavior.]{The list of features used to capture gas pricing behavior, along with their rationale. These features are used to build our models in Section~\ref{sec:rq2} in order to predict transaction (tx) processing times. The $\bullet$ symbol represents data retrieved from Etherscan, while the $\star$ symbol represents data retrieved from Google BigQuery. Braces (\{\}) group a set of related features.}
	\label{tab:add-features}
	\footnotesize
	\renewcommand{\arraystretch}{1.4}
	\resizebox{1.35\textwidth}{!}{%
	\begin{tabular}{p{4.0cm}p{5.8cm}p{5.8cm}}
		\midrule
		\textbf{Name} &
		\textbf{Description} &
		\textbf{Rationale} \\ \midrule
		\texttt{\{avg, med, std\}\_num\_\{above, same, below\}\_120}$\star$ &
		The average, median, and standard deviation of txs. processed per block, in the previous 120 blocks, with a gas price above, equal to, and below the current tx. &
		\multirow{5}{=}{The various aggregations of txs. sent in previous blocks, combined with the aggregations of the associated gas price value, should hold stronger explanatory power than these gas price and processing time features alone. As a result, these features should be powerful in explaining variations in processing times.} 
		\\
		\texttt{\{avg, med, std\}\_pct\_\{above, same, below\}\_120}$\star$ &
		The average, median, and standard deviation of the percentage of txs. processed per block, in the previous 120 blocks, with a gas price above, equal to, and below the current tx. &
		\\
		\texttt{num\_\{above, same, below\}\_\{1, 120\}}$\star$ &
		The total number of txs. processed per block, in the previous 120 blocks and previous block, with a gas price above, equal to, and below the current tx. &
		\\
		\texttt{pct\_\{above, same, below\}\_\{1, 120\}}$\star$ &
		The percentage of the total number of txs. processed per block, in the previous 120 blocks and previous block, with a gas price above, equal to, and below the current tx. &
		\\  \midrule
		\texttt{gas\_price\_cat\_enc}$\star$ &
		The label encoded gas price category of the tx calculated using the percentiles {[}0.2, 0.4, 0.6, 0.8, 1.0{]} of gas prices from all txs. from the previous 120 blocks. &
		\multirow{3}{=}{The gas price is an incentive for miners to prioritize one tx over another. This feature aims to determine whether a certain gas price is \textit{very low}, \textit{low}, \textit{medium}, \textit{high}, or \textit{very high} by comparing it to the prices of recent txs.
		} \\ \\ \midrule
		\texttt{\{avg, med, std\}\_gas\_price\_\{1, 120\}}$\star$ &
		The average, median, and standard deviation of tx gas prices per block, in the previous 120 blocks and previous block & Such aggregations capture current gas prices being paid, and thus should help in predicting future trends in gas prices and its relationship to processing times.
		\\  \midrule
		\texttt{past\_\{avg, med, std\}\_time}$\bullet$ &
		The average, median, and standard deviation of the processing times for the most recent 100 txs. in the previous 120 blocks, with the most similar gas prices as the current tx. &
		\multirow{2}{=}{The aggregations of processing times recorded for the most similar txs. within the previous blocks will give the model a sense of how long the current tx should be taking to be processed.} 
		\\
		\texttt{closest\_tx\_pr\_time}$\bullet$ &
		The processing time of the most recent tx processed in the previous 120 blocks with the most similar gas price as the current tx. &
		\\ \midrule
		&
		&
	\end{tabular}  
	}%
\end{table*}

\subsection{Model Construction}
\label{sec:model-construction}

\begin{figure}[!htbp]
\centering
    \includegraphics[width=1.0\linewidth]{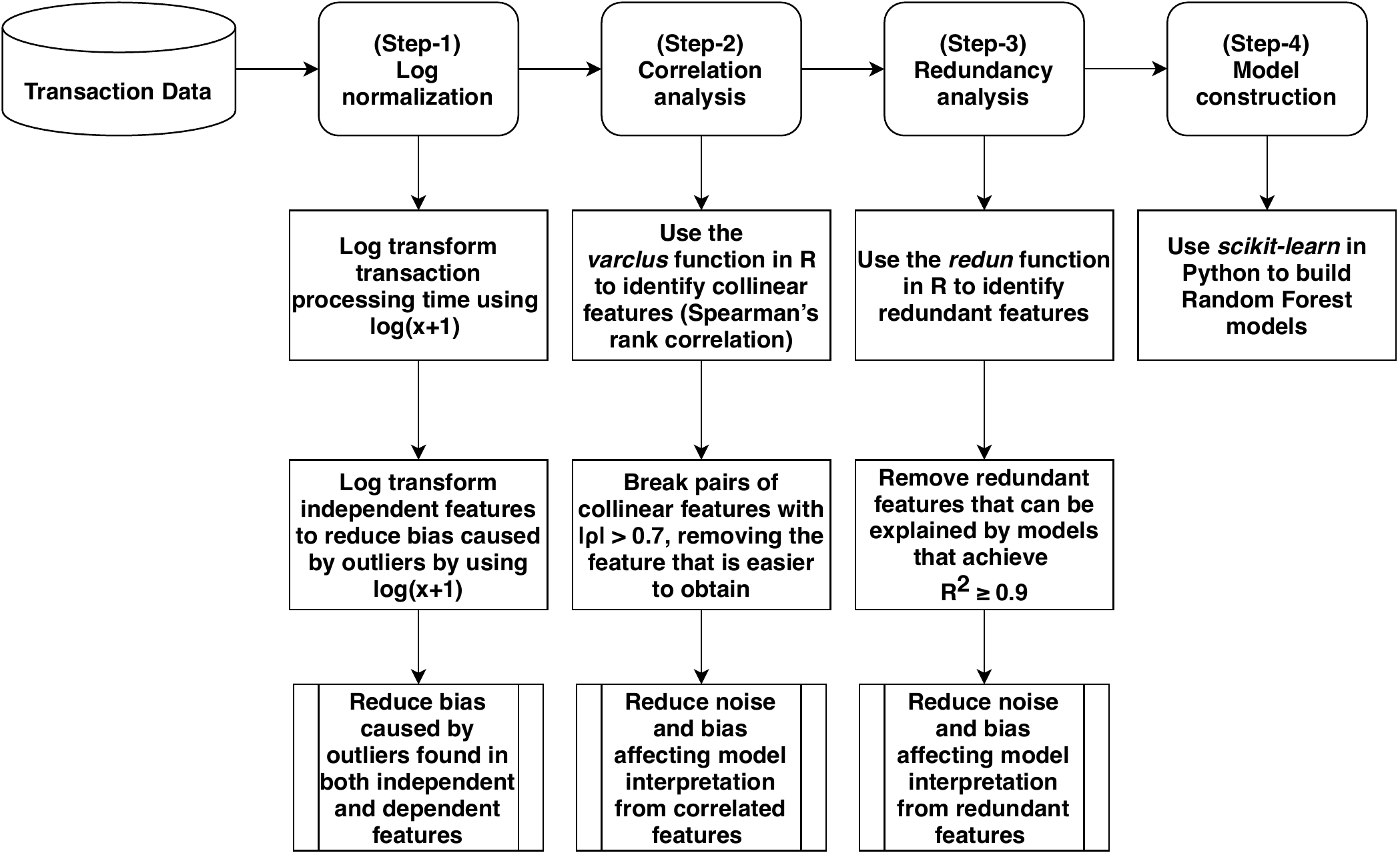}
    \caption{A summary of our model construction approach.}
    \label{fig:model-construction}
\end{figure}

In this section, we describe our model construction steps. The goal of our constructed model is to explain how features related to internal factors and features related to gas pricing behavior influence the transaction processing times respectively. We also aim to answer which of the studied features influence the transaction processing times the most. To do this, we design and follow a generalizable and extensible model construction pipeline to generate interpretable models, as outlined in prior studied \citep{Lee2020, Mcintosh16, Thongtanunam2020}. Rather than focusing on generating models that are meant to be used to predict transaction processing times with high to perfect levels of accuracy, we instead leverage machine learning as a tool to understand the impact of these features. Our model construction pipeline in summarized Figure~\ref{fig:model-construction}. We explain each step in more detail below.


\smallskip \noindent \textbf{(Step-1) Log normalization.} We observe that the distribution of transaction processing time is right skewed. Therefore, we log transform the data with the function $log(x+1)$ to reduce the skew. Furthermore, we also log transform all the independent features of the type \textit{numeric} with an identical function to reduce the bias caused by outliers, similar to prior studies~\citep{Lee2020, Mcintosh16, Menzies2007, tantithamthavorn2017mvt}.

\smallskip \noindent \textbf{(Step-2) Correlation analysis.} Several prior studies show that presence of correlated features in the used data to construct a machine learning model leads to unreliable and spurious insights \citep{Midi2013,Mcintosh16,Kla2018,Lee2020}. Therefore, in this step we outline our process to remove correlated and redundant features in our data. To remove correlated features, we use the \texttt{varclus} function from the \texttt{Hmisc} package in R\footnote{\url{https://www.rdocumentation.org/packages/Hmisc/versions/4.5-0/topics/varclus}}, which generates a hierarchical clustering of the provided features, based on the correlation between them. We choose Spearman's rank correlation ($\rho$) to determine the level of correlation between pairs, as rank correlation can assess both monotonic and non-monotonic relationships. Similar to prior studies~\citep{Lee2020, Mcintosh16, Kla2018}, we use a threshold of $|\rho| > 0.7$ to deem a pair of features as correlated. For each pair of correlated features, we choose the feature that is easier to collect or calculate in practice. We summarize our correlation analysis in Table~\ref{tab:corr_analysis} (Appendix~\ref{appendix:corr_redun_analysis}). As a result of such an analysis, we removed a total of 41 correlated features.

\smallskip \noindent \textbf{(Step-3) Redundancy analysis} In addition to correlated features, redundant features may also negatively impact the generated explanations of a machine learning model~\citep{Midi2013, Mcintosh16, Kla2018, Lee2020}. Hence, we choose to remove redundant features. To do so, we use the \texttt{redun} function from the \texttt{Hmisc} package in R\footnote{\url{https://www.rdocumentation.org/packages/Hmisc/versions/4.5-0/topics/redun}}. This function builds several linear regression models, each of which selects one independent feature to use as the dependent feature. It then calculates how well the selected dependent features can be explained by the independent ones, through the R\textsuperscript{2}. It then drops the most well-explained features in a stepwise, descending order. This feature removal process is repeated until either: 1) none of the features can be predicted by the models resulting in an R\textsuperscript{2} greater than a set threshold, or 2) removing the feature causes another previously removed feature to be explained at a level less than the threshold. In our study, we use the default R\textsuperscript{2} threshold of 0.9. 
As a result of such an analysis, we removed a total of 1 redundant feature, namely \texttt{std\_pct\_below\_120}.

\smallskip \noindent \textbf{(Step-4) Model construction} At this step, we fit our random forest model using the \texttt{RandomForestRegressor} function from the \texttt{scikit-learn} package in Python\footnote{\url{https://scikit-learn.org/stable/modules/generated/sklearn.ensemble.RandomForestRegressor.html}}. 
We choose a random forest model as it provides an optimal balance between predictive power and explainability. Such models are equipped to fit data better than simple regression models, potentially leading to more robust and reliable insights \cite{Yingzhe2021}. Although they are not inherently explainable, several feature attribution methods can and have been used to derive explanations using these models~\cite{Breiman01, Lundberg18, tantithamthavorn_xai4se}. Random forests have also been widely used in several empirical software engineering research to derive insights about the data~\cite{Gopi21, tantithamthavorn_xai4se, tantithamthavorn2020}. Historically these models have shown good performance results for Software Engineering tasks~\citep{Yatish19, Bao19, Fan20, Aniche21}. Finally, the specific random forest implementation that we choose is natively compatible with the \texttt{shap}\footnote{\url{https://shap.readthedocs.io/en/latest/index.html}} python package. The latter implements the SHAP interpretation technique, which we employ to address our research questions (Section~\ref{sec:rq1} and Section~\ref{sec:rq2}).

\subsection{Model validation}
\label{sec:model-validation}

\begin{figure}[!htbp]
\centering
    \includegraphics[width=0.8\linewidth]{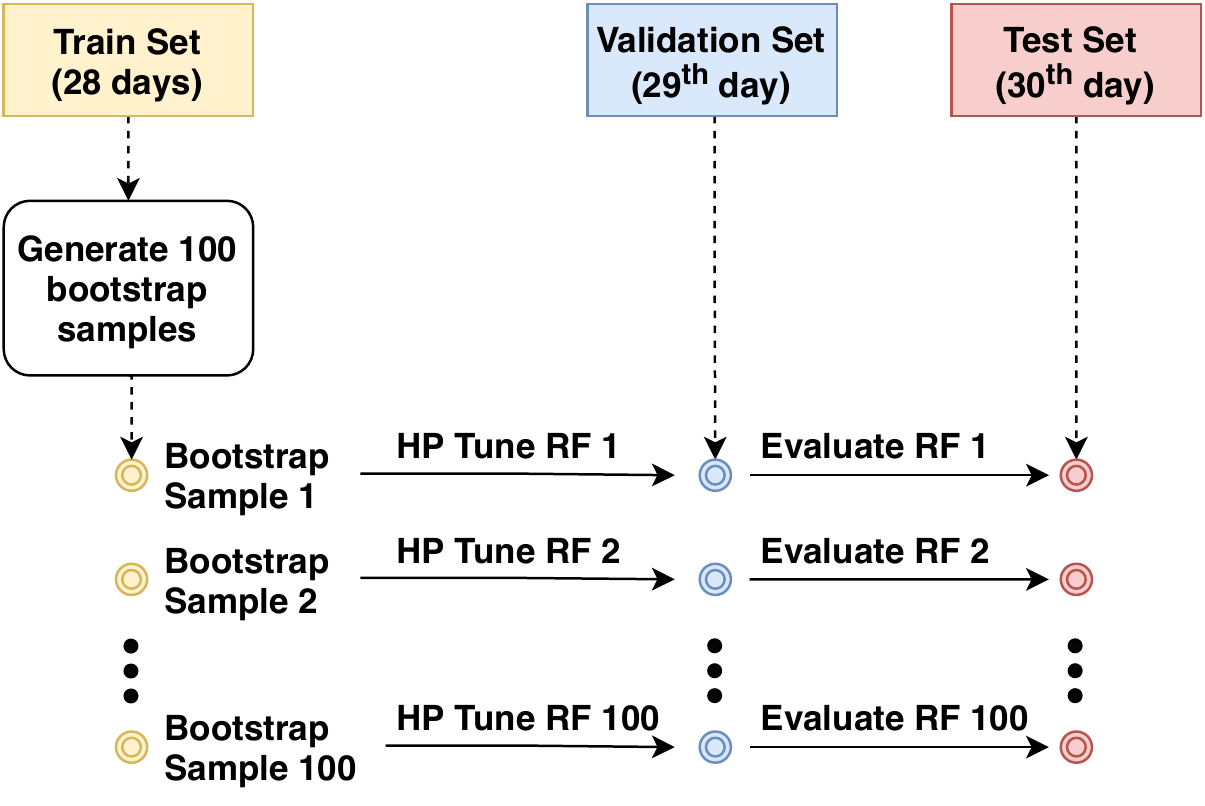}
    \caption{Our model validation approach. RF stands for Random Forest and HP stands for hyperparameter.}
    \label{fig:model-validation}
\end{figure}

We use the adjusted $R^2$ to evaluate how well our models can predict transaction processing times. To compute the adjusted $R^2$ of our constructed random forest model, we set up a model validation pipeline that splits data into train, test, and validation sets (Figure~\ref{fig:model-validation}). Our train dataset needs to be as large as possible, otherwise our model cannot identify possible seasonalities in the data (e.g., evaluate the influence of the \textit{day of the week} in transaction processing times) nor changes in the blockchain context (e.g., network conditions are unlikely to change in small periods of time). We thus define the training set as the first 28 days of data, the validation set as the 29\textsuperscript{th} day, and the test set as the final day. We split the data along its temporal nature to avoid data leakage~\cite{Yingzhe2021}. We then use the training set to generate 100 bootstrap samples (i.e., samples with replacement and of the same length as the training set). For each bootstrap sample, we first fit a random forest model on that sample. Next, we use the validation set to hyperparameter tune that model as recommended by prior studies to ensure that our constructed model fits the data optimally. To do this we utilize a Random Search algorithm, and choose the \texttt{RandomizedSearchCV} \citep{Bergstra2012} from \texttt{scikit-learn}\footnote{\url{https://scikit-learn.org/stable/modules/generated/sklearn.model_selection.RandomizedSearchCV.html}}. We choose to hyperparameter tune the parameters: \texttt{max\_depth}, \texttt{max\_features}, and \texttt{n\_estimators}. Once we discover the optimal hyperparameters, we retrain the model using the chosen parameters (on the same bootstrap sample) and evaluate it on the test set by adjusted $R^2$ measure. While prior studies typically measure the explanatory power of a constructed model by measuring its $R^2$ score on the training set, we instead measure the performance of our hyperparameter-tuned model on a hold-out test set. Our reasoning is that random forest models are by default generally constructed in a way which causes all training data points to be found in at least one terminal node of the forest. As a consequence, it is common for the random forest to predict correctly for the majority of instances of the train set, resulting in an R2 that is very close to 1 (typically in the 0.95 to 1.0 range) \cite{Liaw}. In addition, we use \textit{adjusted} version of the $R^2$ measure, which mitigates the bias introduced by the high number of features used in our model~\citep{Harrell15}. As we generate 100 bootstrap samples, we obtain 100 adjusted $R^2$ values. We report summary statistics for this distribution of adjusted $R^2$ values in the research questions below.

\begin{small}
    \begin{mybox}{Summary}
        \begin{itemize}[itemsep = 3pt, label=\textbullet, wide = 0pt]
            \item Total collected transactions: 1.8M transactions, spanning over 10.8k blocks.
            \item Data sources: Etherscan and Google BigQuery.
            \item Pieces of collected data: Metadata for blocks (e.g., number of included transactions), transactions (e.g., gas price parameters, processing times), contracts (e.g., whether it implements the ERC20 interface), network utilization over time, pending pool size over time.
            \item Raw feature-set (pre-filtering): 75 features (9 contextual, 5 behavioral, 20 historical, and 41 for gas pricing behavior)
            \item Final feature-set (post-filtering): 30 features (6 contextual, 4 behavioral, 9 historical, and 11 for gas pricing behavior).
            \item Machine learning algorithm: Random forest regressor.
        \end{itemize}
    \end{mybox}
\end{small}

	\section{RQ1: \RQOne}
\label{sec:rq1}

\newcommand{\rawftone} {gas\_price\_gwei\xspace}
\newcommand{\rawfttwo} {med\_pend\_prices\xspace}
\newcommand{\rawftthree}{tx\_nonce\xspace}

\noindent \textbf{Motivation.} It is currently unknown which factors from the internal workings of the Ethereum blockchain best characterize transaction processing times. There exists an exhaustive set of information recorded on the blockchain at any given moment in time. For \dapp developers, it is unclear which of these factors are useful in predicting processing times, and thus which should be monitored and analyzed.

\smallskip \noindent \textbf{Approach.} We analyze the model's explanatory power and feature importances. The details are shown below:

\smallskip \noindent \textit{-- Analysis of a model's explanatory power.} To determine the explanatory power of blockchain factors, we engineer features based on the \textit{internal} data recorded and available in the Ethereum blockchain (Table~\ref{tab:features}). Next, we use these feature to construct (Section~\ref{sec:model-construction}) and validate (Section~\ref{sec:model-validation}) our random forest model. As a result of applying our model validation approach, we obtain a distribution of 100 adjusted $R^2$ scores. This score distribution determines how much of the total variability in processing times is explained by our random forest model, thus serving as a good proxy for the explanatory power of our model. We judge the explanatory power of our model by analyzing the mean and median adjusted $R^2$ scores.

\smallskip \noindent \textit{-- Analysis of feature importance.} To determine what individual features are most strongly associated with processing times, we compute and analyze feature importance using SHAP~\citet{Lundberg17, Molnar19}. SHAP is a model-agnostic interpretation technique based on the game theoretically optimal \textit{Shapley values}. SHAP first determines an average prediction value known as the \textit{base rate}. Next, for each dataset instance, SHAP determines the Shapley value of each feature. The Shapley value informs the degree to which each feature moves the base rate (positively or negatively), in the same scale as the dependent variable. The final prediction is explained as: \textit{base rate} + $\sum_{i = 1}^{n} Shapley(f_i)$, where $f_i$ is one of the \textit{n} features of the model. The importance of a feature $f_i$ corresponds to its distribution of Shapley values across all dataset instances. More details about SHAP can be found in Appendix~\ref{appendix:shap}.

To compute Shapley values, we first train a single random forest model on the entire \textit{train set} (28 consecutive days) using the procedures that are outlined in Section~\ref{sec:model-construction}. Next, we use the \texttt{TreeExplainer} API from the \texttt{shap} python package to compute the Shapley value of each feature for each dataset instance. As a result, each feature will have an associated Shapley value distribution of size $N$, where $N$ is the total number of rows in our train dataset. Next, we convert these Shapley value distributions into \textit{absolute} Shapley value distributions. The rationale of such a conversion is that a feature $f_1$ with a Shapley value of $x$ is as important as a feature $f_2$ with a Shapley value of $-x$. Finally, we employ Scott Knott algorithm to rank the distributions of absolute Shapley values. The feature ranked \textit{first} is the one with the highest Shapley values. We note that there can be a rank tie between two or more features, which indicates that their underlying shapley distributions are indistinguishable from an effect-size standpoint. We also note that we use Menzies' implementation~\citep{TimSK} of the original SK algorithm~\citep{Scott74}, which (i) replaces the original parametric statistical tests with non-parametric ones and (ii) takes effect size into consideration (Cliff's Delta). Menzies' changes make the SK algorithm suitable to be used in the context of large datasets, where p-values are close to meaningless without an accompanying effect size measure and few or even no assumptions can be made in terms of the shape of the distributions under evaluation.

\smallskip \noindent \textbf{Findings.} \observation{Internal factors provide little explanatory power in the characterization of transaction processing times.} 
Our random forest models achieve identical mean and median adjusted R\textsuperscript{2} scores: 0.16. These scores indicate that internal factors can only explain approximately 16\% of the variance in processing times.

\begin{figure}[!htbp]
    \centering
    \includegraphics[width=1.0\linewidth]{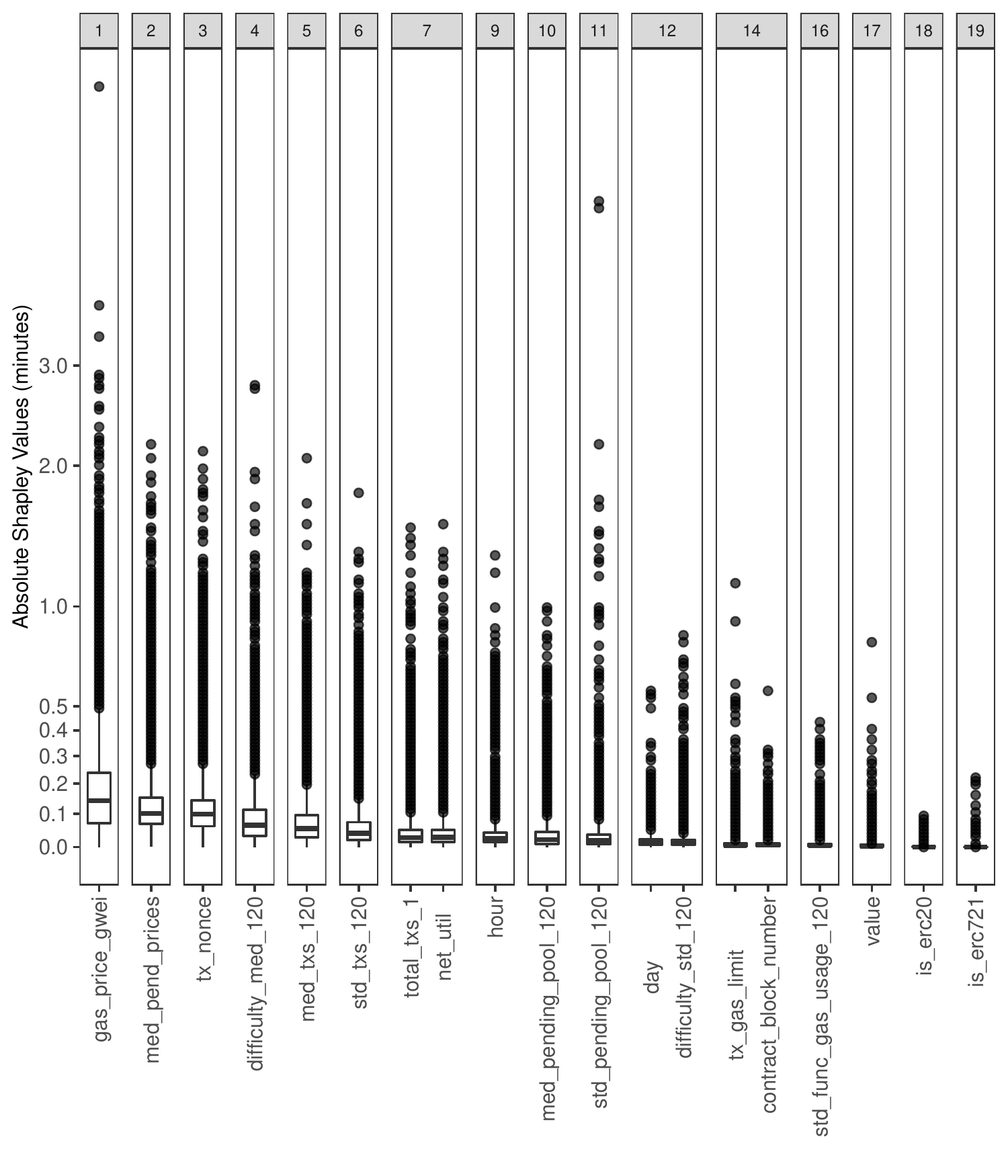}
    \caption{The distribution of absolute Shapley values per feature (minutes). Features are grouped according to their importance rank (lower rank, higher importance). Due to the high skewness of the data, we apply a log(x+1) transformation to the y-axis. 
    }
    \label{fig:raw-shap-dist}
\end{figure}

\smallskip \observation{Gas price, median price of pending transactions, and nonce are the top-3 most important features of the model.} Feature importance ranks are summarized in Figure~\ref{fig:raw-shap-dist}. It is unsurprising that the \textit{transaction's gas price} (\texttt{\rawftone}) is the most important feature, since the incentive that miners receive to process transactions is a function of the gas price of the transactions that they choose to process. Nevertheless, we reiterate that our model achieves a median adjusted $R^2$ of only 0.16  (Observation 1). That is, despite its inherent importance, the gas price of a transaction explains less than 0.16 of the variability in transaction processing times. We conjecture that the explanation lies in the \textit{relative} value of a given gas price. In a way, gas prices can be interpreted as bids in an auction system. The actual, practical value of a bid will always depend on the bids offered by others. Hence, it is entirely possible that the same gas price of x GWEI could be regarded as \textit{high} in one day (i.e., when other transactions typically have a gas price that is lower than x) and \textit{low} in some other day (i.e., when the opposite scenario takes place). We explore this conjecture as part of RQ2, when we evaluate the explanatory power of pricing behavior.

The second most important feature is the \textit{median gas price of recent pending transactions from the same transaction issuer} (\texttt{\rawfttwo}). In Ethereum, a transaction $t$ from a given transaction issuer $I$ can only be processed once all prior transactions from $I$ have been processed (i.e., those transactions with $nonce$ lower than that of $t$). For instance, a transaction can stay in a pending state for a long time when a prior transaction from the same issuer has a very low gas price. Hence, pending transactions penalize the processing time of the current transaction. In fact, if the value of \texttt{\rawfttwo} is higher than zero, then we know by construction that there exists at least one pending transaction (since gas price cannot be zero). We assume that our model has learned this rule. Interestingly, the third most important feature is the \textit{transaction's nonce} (\texttt{\rawftthree}). Hence, we believe that our model is combining the second and third most important features to estimate the time penalty induced by pending transactions. 

There is also a complementary facet to \texttt{\rawftthree}, which is bound to the transaction issuer. A transaction with a high nonce indicates that its transaction issuer has submitted many transactions in the past. Experienced transaction issuers are likely to be more careful in setting up their gas prices in comparison to novice issuers (e.g., they might be more mindful about the competitiveness of their chosen gas prices compared to novices)~\citep{Oliva21}. 


\smallskip \observation{Contextual features have little importance.} The most important contextual feature is ranked only 7th. Its median Shapley value is 0.03 minutes (1.8 seconds). In other words, in at least 50\% of the cases, this features either adds or removes only 1.8 seconds to the final prediction.


\begin{small}
    \begin{mybox}{Summary}
        \textbf{\RQOne}
        \tcblower
        Internal factors provide little explanatory power in the characterization of transaction processing times. In particular:
        
        \begin{itemize}[topsep = 3pt, itemsep = 3pt, label=\textbullet, wide = 0pt]
        
            \item Our random forest models only achieve both mean and median adjusted $R^2$ scores of 0.16.
            \item Gas price, median price of recent pending transactions from the same transaction issuer, and nonce are the top-3 most important features of our model.
            \item Contextual features have little importance.
            
        \end{itemize}
    \end{mybox}
\end{small}
	\section{RQ2: \RQTwo}
\label{sec:rq2}

\newcommand{\fullftone} {med\_pct\_below\_120\xspace}
\newcommand{\fullfttwo} {med\_pend\_prices\xspace}
\newcommand{\fullftthree}{std\_num\_below\_120\xspace}

\noindent \textbf{Motivation.} The results of our previous research question indicates that internal factors of the Ethereum blockchain only explain a small amount of the variability in transaction processing times. 

Similar to the stock market, the Ethereum blockchain ecosystem is impacted by a plethora of external factors, including speculation (e.g., NFTs, Elon Musk Tweets), the global geopolitical situation (e.g., wars, the USA-China trade war), cryptocurrency regulations (e.g., FED interventions), and even unforeseen events (e.g., the COVID-19 pandemic). Many of these factors are largely unpredictable and difficult to be captured in the form of engineered features. Yet, these factors are known to impact the \textit{gas pricing behavior} of a large portion of transaction issuers~\cite{Ante21, Mashable}. For instance, a cryptocurrency ban in a country can cause a massive amount of people to sell their ETH and tokens, causing many \dapps to send transactions to facilitate those actions, thus increasing the average gas price being paid for all transactions in the platform. Since the practical value of a gas price depends on the gas price of all other transactions, we conjecture that pricing behavior (and changes in thereof) influences the processing times of transactions.

\smallskip \noindent \textbf{Approach.} The additional features that we engineer in order to capture gas pricing behaviors are described in Table \ref{tab:add-features} along with their rationale. In summary, we introduce 42 additional features to our internal feature set. We primarily involve aggregations of the number of previously processed transactions in recent blocks and their processing times (historical features), in conjunction with the gas price of these transactions (a behavioral feature). For instance, one of our features indicates whether a given gas price is very low/low/median/high/very high by comparing it to the gas prices of all transactions included in the prior 120 mined blocks. In a way, our model from RQ2 provides a \textit{relative} perspective on gas prices instead of an absolute one (RQ1).


Our approach is similar to that of RQ1. First, we reuse the model validation approach from Section~\ref{fig:model-validation} to determine the explanatory power of our new model (adjusted $R^2$). Next, we compare the difference in adjusted $R^2$ between the models from RQ1 and RQ2. The rationale is to determine the importance of the additional features (pricing behavior). We operationalize the comparison by means of a two-tailed Mann-Whitney test ($\alpha = 0.05$) followed by a Cliff's Delta (d) calculation of effect size.  We evaluate Cliff's Delta using the following thresholds~\citep{Romano06}: negligible for $|d| \leq$ 0.147, small for 0.147 $\le |d| \leq$ 0.33, medium for 0.33 $\le |d| \leq$ 0.474, and large otherwise. Finally, to understand the importance of individual features, we plot their distribution of absolute Shapley values and compute SK.

To further understand the role of individual features, we use \textit{Partial Dependence Plots} (PDP). A PDP reflects the expected output of a model if we were to intervene and modify exactly one of the features. This is different from SHAP, where the Shapley value of a feature represents the extent to which that feature impacts the prediction of single dataset instance while accounting for interaction effects. We also compute pair-wise feature interactions to understand the interplay between specific pairs of features. 
We use the \texttt{explainer.shap\_interaction\_values()} function from the same \texttt{shap} package. 
This computes the Shapely interaction index after computing individual effects, and does so by subtracting the main effect of the features to result in the pure interaction effects~\cite{Molnar19}.

\smallskip \noindent \textbf{Findings.} \observation{Our new model explains more than half of the variance in processing times.} With the new feature set, our random forest models achieve identical mean and median adjusted R\textsuperscript{2} scores: 0.53. In RQ1, the mean and median scores were also identical at 0.16. As the numbers indicate, the pricing behavior dimension leads to an absolute increase of 0.37 in the median/mean adjusted R\textsuperscript{2}. Indeed, the difference between the adjusted $R^2$ distributions is statistically significant (p-value $\leq$ 0.05) with a large (d = 1) effect size. We find these scores confirm the robust explanatory power of our features, demonstrating the stability of our models. 
In particular, they are compatible (and even sometimes outperform) adjusted $R^2$ scores reported in software engineering studies that built regression models with the purpose of understanding a certain phenomenon \cite{Bird2011, Jiarpakdee2021, Mcintosh16}. For instance, \citet{Mcintosh16} built sophisticated regression models using restricted cubic splines to determine whether there is a relationship between (i) code review coverage and post-release defects, (ii) code review participation and post-release defects, and (iii) reviewer expertise and post-release defects. In total, the authors built 10 models and obtained adjusted $R^2$ scores that ranged from 0.20 to 0.67 (mean = 0.46, median = 0.41, sd = 0.17). In comparison, the median adjusted $R^2$ score of our model is 0.53, which sits higher than the median score achieved by their models. Most importantly, our score was computed on a \textit{hold-out test set}, while theirs were computed on the \textit{train set}. As the hold-out test set is unseen by the model during training, it is generally harder to achieve a higher $R^2$ on compared to the train set.

\begin{figure}[!htbp]
    \centering
    \includegraphics[width=1.0\linewidth]{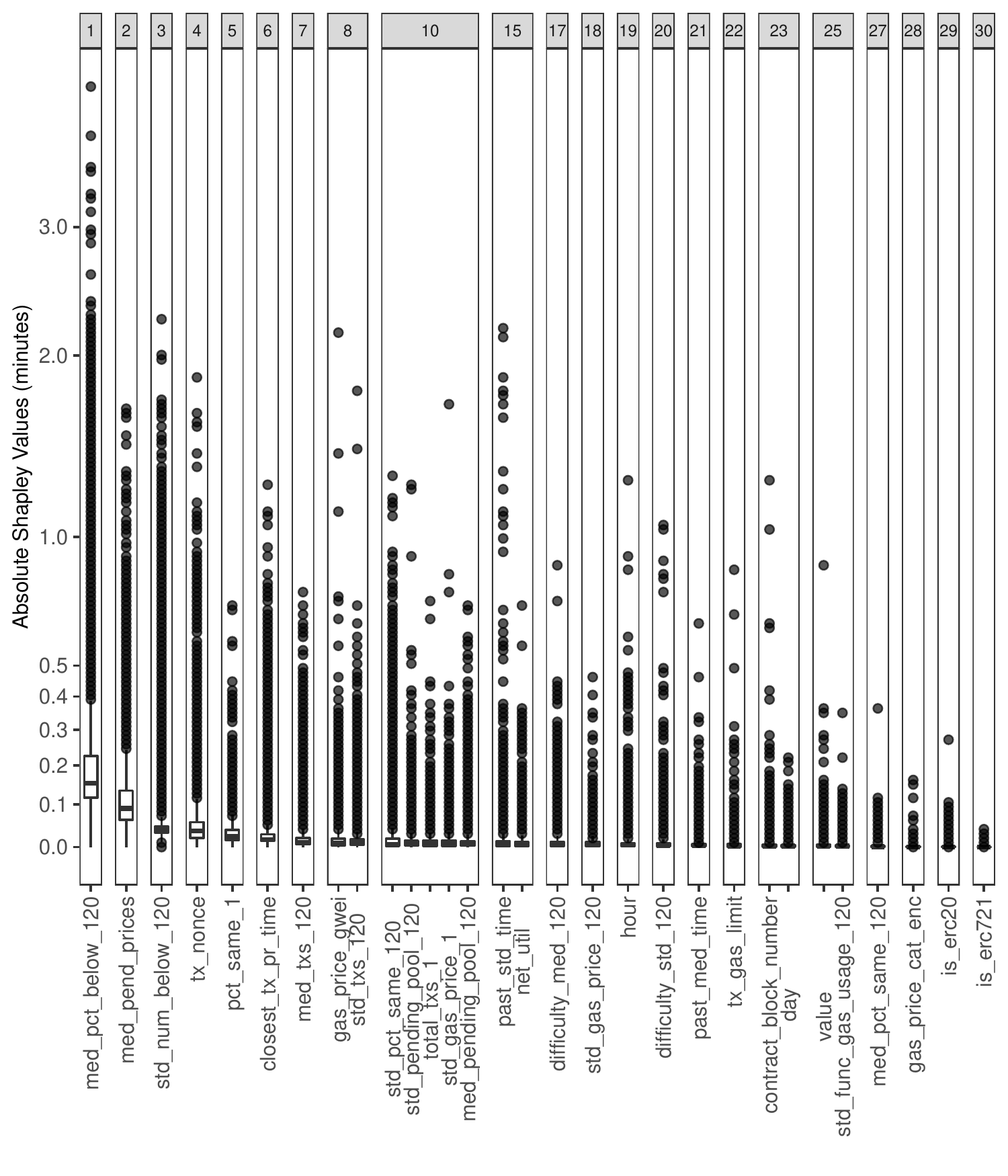}
    \caption{The distribution of absolute Shapley values per feature in our model with additional features. Features are grouped according to their importance rank. Due to the high skewness of the data, we apply a log(x+1) transformation to the y-axis. 
    }
    \label{fig:full-shap-dist}
\end{figure}

\smallskip \observation{Gas price competitiveness is a major aspect in determining processing times.} Figure~\ref{fig:full-shap-dist} depicts the distribution of absolute Shapley values associated with each feature. The most important feature is the \textit{median of the percentage of transactions processed per block, in the previous 120 blocks, with a gas price below that of the current transaction} (\texttt{med\_pct\_below\_120}). The gas price per se (\texttt{gas\_price\_gwei}) only ranks 8th, with a small median absolute Shapley value of 0.01. 


\begin{figure}[!htbp]
    \centering
    \includegraphics[width=\linewidth]{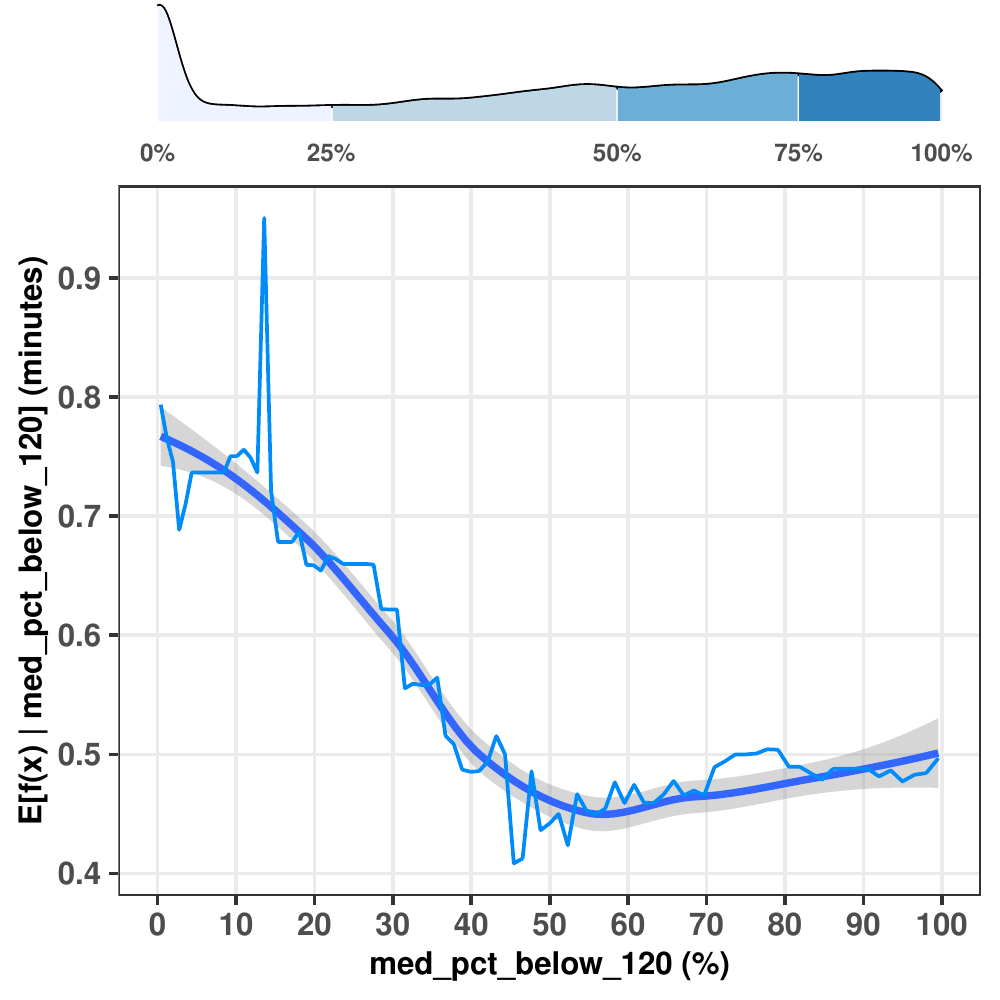}
    \caption{A partial dependence plot for the \texttt{med\_pct\_below\_120} feature and transaction processing time.}
    \label{fig:full-pdp-1}
\end{figure}

\smallskip \observation{There is a gas price at which transactions are processed closest to the fastest speed possible (and thus increasing such a price is *not* likely to further reduce the processing time.)} The PDP depicted in Figure~\ref{fig:full-pdp-1} leads to several insights. First, the range of the y-axis indicates that changes in \texttt{med\_pct\_below\_120} have a considerable influence on the predicted processing times. Second, we note an inverse relationship between \texttt{med\_pct\_below\_120} and the predicted processing time up until \texttt{med\_pct\_below\_120 = 50\%}. That is, let $p(B)$ be the lowest gas price used by a transaction inside a mined block B. Let $p(t)$ be the gas price of a transaction $t$. Setting $p(t)$ such that $p(t) > p(B)$ is true for at least half of the 120 most recently mined blocks already provides a speed boost in transaction processing time. From \texttt{med\_pct\_below\_120 = 50\%} onwards, however, the processing time does not decrease anymore. 

In summary, (i) increasing the gas price beyond a certain value $x$ does not tend to reduce processing times any further and (ii) the value of $x$ is dependent on the prices of the transaction in the prior 120 blocks (otherwise the \texttt{gas\_price\_gwei} would be more important than \texttt{med\_pct\_below\_120}).

\smallskip \observation{Information about the median price of pending transactions and nonce remain important factors.} Looking at Figure~\ref{fig:full-shap-dist} again, we note that the feature \textit{median price of pending transactions from the same issuer} (\texttt{med\_pend\_prices}) and \textit{transaction nonce} (\texttt{tx\_nonce}) are ranked second and fourth. In RQ1, our baseline model suggests that these same two features were ranked second (the same) and third (one rank lower) respectively. In other words, these two features remain important factors.

We plotted the PDP for \textit{median price of pending transactions from the same issuer} (\texttt{med\_pend\_prices}) and observed that the y-axis range is very constrained (range of 0.01 for 90\% of the values of \texttt{med\_pend\_prices}). In other words, the predicted processing time changes very little when \texttt{med\_pend\_prices} is changed and everything else kept constant. Nevertheless, such a feature ranks second according to SHAP (Figure~\ref{fig:full-shap-dist}). Hence, we conjecture that \texttt{med\_pend\_prices} has strong interactions with some other feature. In fact, in RQ1 we conjectured that the model could be using both \texttt{med\_pend\_prices} and \texttt{tx\_nonce} to penalize processing time due to pending transactions. To investigate our conjecture, we calculated feature interactions between \texttt{med\_pend\_prices} and every other feature. The results indicate that the strongest interaction is indeed with \texttt{tx\_nonce}. We then computed all pairwise feature interactions and noted that the feature pair <\texttt{med\_pend\_prices}, \texttt{tx\_nonce}> is the second-strongest feature interaction in our model, only behind <\texttt{med\_pct\_below\_120}, \texttt{std\_num\_below\_120}> (i.e., the first and third most important features).

\begin{small}
    \begin{mybox}{Summary}
        \textbf{RQ2: \RQTwo}
        \tcblower
        Gas pricing behavior characterize transaction processing times much better than blockchain internal factors alone. In particular:

        \begin{itemize}[topsep = 3pt, itemsep = 3pt, label=\textbullet, wide = 0pt]
            \item Our random forest models achieve a median adjusted R\textsuperscript{2} scores of 0.53. 
            \item Gas price competitiveness is a major aspect in determining processing times.
            \item Setting a gas price above a certain value is unlikely to further reduce processing time.
            \item Number and gas price of pending transactions have a large influence on the processing time of the current transaction.
        \end{itemize}
    \end{mybox}
\end{small}
	\section{Discussion}
\label{sec:discussion}

In the following section, we discuss the implication drawn from our findings in the previous sections.

\smallskip \implication{\dapp developers should focus on quantifying how competitive their gas prices are.}Price competitiveness is the most important indicator for transaction processing times among our entire set of engineered features. In particular, it is significantly more important than internal blockchain factors that practitioners typically associate with transaction processing times, including: block difficulty (ranks 17 and 20), the number of transactions in the pending pool (rank 10), and network utilization (rank 15).

In practice, \dapp developers should ensure that they execute transactions with gas competitive gas prices and the competitiveness of a given specific price is likely to change over time. Given our empirical results, we believe that our metric \textit{med\_pct\_below\_120} can serve as a starting point for \dapp developers to quantify price competitiveness. We warn developers, however, that such a price should not be ``too high'', since there is a point at which higher prices do not result in lower processing times. More generally, future research in explainable models for transaction processing times should investigate the optimal number of past blocks to consider and how frequently such a number needs to be fine-tuned over time to avoid concept drift.

\smallskip \implication{The architecture of \dapps should take into account the fact that processing time is only weakly associated with gas usage and gas limit.} The contribution of work lies in not only identifying those feature that are most important, but also in identifying those that are not. 

Figure~\ref{fig:full-shap-dist} indicates that features related to \textit{gas usage}, such as \textit{tx\_gas\_limit} and \textit{std\_func\_gas\_usage\_120} (which is correlated with \texttt{med\_func\_gas\_usage\_120}), have little importance. This suggests that transactions with competitive gas prices are prioritized by miners over the amount of gas that they will potentially consume. Hence, if a \dapp invokes a heavy computation function that consume lots of gas, then such a transaction will still require a competitive gas price to be processed in a timely fashion. Such a \dapp might be costly and potentially unfeasible to maintain. A more frugal approach would be to design a \dapp architecture in which heavy computations are accomplished by means of multiple transactions that consume little gas each. We acknowledge, however, that such an architecture might be unfeasible in specific scenarios or use cases (e.g., when the computation needs to be performed as soon as possible). Most importantly, our work highlights an important and yet not so obvious interplay between \dapp architecture design and transaction processing times.

\smallskip \implication{We did not observe a strong relationship between contextual features and processing time. However, future research should not discard this type of feature.} We observed that features in the contextual dimensions generally contribute very little to the resulting prediction. This may occur either because (i) contextual really do not matter that much compared to pricing behavior or (ii) because contextual factors are not strongly present in our dataset (e.g., seasonalities and network congestion). We thus invite future research to further explore contextual features with other datasets and timeframes, as well as engineer additional contextual features.
	\section{Related Work}
\label{sec:related-work}
We note that the analysis we conduct in our study on the explanatory power of an exhaustive set of features with respect to transaction processing times has not yet been investigated. We discover several insights using this proposed approach, the majority of which suggest the importance of historical features related to gas prices. We hope our contributions prove useful for \dapp developers, and invite future research to build upon our study. 

\citet{Pierro19} investigate how several factors related to transaction processing times cause or correlate with variations in gas price predictions made by Etherchain’s Gas Price oracle\footnote{\url{https://etherchain.org/tools/gasPriceOracle}}. 
The authors conclude there is non-directional causality between oracle gas price predictions and the time to mine blocks, a unidirectional causality (inverse correlation) between oracle predictions and number transactions in the pending pool, and a unidirectional causality (inverse correlation) between oracle predictions and the number of active miners.

In a later study the same authors \citet{Pierro21} evaluate the processing time and corresponding gas price predictions made by EthGasStation's Gas Price API. To do this, the authors use processed transactions by retrieving those with a gas price greater or equal to the gas prices in all processing time predictions made by EthGasStation. They then verify whether those transactions were processed in the following $j$ blocks, as predicted by EthGasStation. The authors conclude that EthGasStation holds a higher margin of error compared to what they claim. Using a Poisson model, the authors achieve more accurate predictions than EthGasStation by considering data within only the previous 4 blocks. In turn, the authors suggest that such models should be constructed using data within the most recent blocks, as opposed to the 100 previous blocks considered by EthGasStation, to improve prediction accuracy.

\citet{deAzevedo21} investigate the correlation between transaction fees and transaction pending time in the Ethereum blockchain. 
The authors calculate Pearson correlation matrices between several features across 1) the full set of transactions, 2) distinct sets of transactions based on their gas usage and gas price, and 3) clusters of transactions resulting from DBSCAN. The authors conclude there is a negligible correlation between all features related to transaction fees and the pending time of transactions.

\citet{Singh20} focus on the prediction of transaction confirmation time within the Ethereum blockchain. The study includes 1 million transactions with non-engineered features from the blockchain. The authors discretize transactions by their confirmation times (15s, 30s, 1m, 2m, 5m, 10m, 15m, and $\geq$ 30m).
The majority of constructed classifiers (Naive Bayes, Random Forests, and Multi Layer Perceptrons) achieved higher than 90\% accuracy, with the Multi Layer Perceptron classifier regularly outperforming the others.
The impact of the transaction fee mechanism (TFM) EIP-1559 on processing times is investigated by \citet{liu2022} in the Ethereum blockchain since its relatively recent implementation. 
The authors discover that EIP-1559 makes fee estimation easier for users, ultimately reducing both transaction processing times and gas price paid of transactions within the same block. However, when the price of Ether is volatile, processing times become much longer. We find that features related to historical gas prices are most important in estimating processing time, and considering EIP-1559 makes fee estimations and thus assigning gas prices easier for users, similar TFMs may help \dapp developers avoid slow processing times altogether, 

The study conducted by \citet{Kasahara} uses queuing theory to investigate the impact of transaction fees on transaction processing time within the Bitcoin blockchain. The authors conclude that transactions associated with high transaction fees are processed quicker than those that are not. The authors also find network congestion strongly impacts the processing time of transactions, by increasing the time, regardless if they are associated with either low or high fees, and regardless of block size. Conversely, our study finds network utilization to hold minimal impact on transaction processing time.


\citet{Oliva} analyze how smart contracts deployed on the Ethereum blockchain are commonly used. The authors conclude that 0.05\% of total smart contracts were the target of 80\% of all transactions executed during the time period of their study. In addition, 41.3\% of these contracts were found to be deployed for token-related purposes. Our paper includes information regarding the smart contract targeted by a transaction if applicable, though we do not find that these features have strong impacts on transaction processing time.

The work of \citet{Zarir21} investigates how \dapp developers can develop cost effective \dapps. Similar to our study, the authors conclude transactions are prioritized based on their gas price, rather than their gas usage. In addition, the authors find the gas usage of a function can be easily predicted. As a result, the authors suggest that \dapp developers provide gas usage information in their smart contracts, and platforms such as Etherscan and wallets should provide users with historical gas usage information for smart contract functions.
	\section{Threats to Validity}
\label{sec:threats}


\noindent \textbf{Construct Validity}. The \textit{pending timestamp} of a transaction is not publicly available information within the Ethereum blockchain. As a result, we rely on Etherscan to provide an accurate measurement of when a transaction was actually executed. Because of the size, reputability, and popularity of Etherscan, we continue to rely on it and reaffirm the accuracy of their provided data.

This paper is a first attempt toward using predictive models in order to derive and explain insights from data within the context of the Ethereum blockchain. We choose to leverage Random Forest models as they hold an optimal balance between performance and explainability. We encourage future work to build upon the approach demonstrated in our study in order to achieve models which achieve higher performance, and thus further contribute to predicting transaction processing times and the generalizability of the model explanations in our study.

Our Random Forest models are hyperparameter tuned in each bootstrap to control states of the model as well as to achieve greater performance on the hold out test set. We only consider a few hyperparameters to control, and there exist multiple hyperparameters that can be tuned which were not considered in our study. We invite future researchers to consider a more exhaustive set of hyperparameters to tune in similar Random Forest models, and to investigate their impact on performance.

Although we opt to leverage the adjusted $R^2$ metric as a robust approach to determine the goodness of fit of the model, other performance metrics exist that can, in certain contexts, contribute toward a complete assessment of the performance of the model. As such, future work should assess similar models using additional performance metrics.

We fully rely on SHAP values to generate explanations for the predictions of our Random Forest models. It is possible that different feature importance methods would generate different results. We thus encourage future work to explore other methods for explaining models (e.g., LIME~\cite{Ribeiro2016}).

\smallskip \noindent \textbf{Internal Validity}. In our paper, we choose to study the explanatory power of a set of both internal and external features spanning across various internal dimensions. The majority of features in our set include those which primarily reflect technical aspects of the Ethereum blockchain. However, other types of features, such as those that are based upon and or track market speculation (e.g., Elon Musk tweets \cite{Ante21}), exist which might play a big role in describing the behavior of the Ethereum blockchain. As a result, we invite future work to design and investigate an even more exhaustive set of features which were not included in our study.

The collected data and concluding results naturally depend on the defined time frame of our study. We also use sampling methods to construct our results, though we draw a representative sample of blocks appended to the Ethereum blockchain per day, in attempt to capture the complete and natural behavior of the network during this period. Seasonal trends are known to impact the blockchain \cite{Kaiser2019, Werner2020} and therefore may have affected the insights and conclusions derived in our study. As a result, we suggest that future studies should validate our findings by replicating it using different and longer time frames. Finally, we emphasize that in this paper we demonstrate an approach that is extensible, and therefore can be followed in similar contexts with similar goals.


In addition to the dependence on the chosen time frame, our study also relies on actively collecting data from Etherscan. As Etherscan does not offer a robust API which allows us to collect all data points used in our study, we developed other methods of data collection. Our data collection methods respect Etherscan's platform by adhering to standards they have set so as to not disrupt the availability of their normal services. Because of this, we are unable to collect every transaction executed within the Ethereum blockchain, which might result in imbalanced data, such as collecting a low amount of transactions that are processed in extremely fast or slow times. 

\smallskip \noindent \textbf{External Validity}. Our study focuses solely on the Ethereum blockchain during a specific time frame. As it is likely for blockchains to be unique in purpose and design, the consistency of the conclusions resulting from our study may not hold in other blockchain environments. We encourage future research to explore ideas and conduct replication experiments similar to ours in other blockchain environments, as well as different time frames, which may be fruitful.
	\section{Conclusion}
\label{sec:conclusion}



\dapps on the blockchain require code to be executed through the use of transactions, which need to be paid for. For \dapps to be profitable, developers need to balance paying high amounts of Ether to have their application transactions processed timely, and high-end user experience. Existing processing time estimation services aim to solve this problem, however they offer minimal insight into what features truly impact processing times, as the platforms to not offer any interpretable information regarding their models.

With this as motivation, we collect data from Etherscan, and Google BigQuery to engineer and investigate features which capture internal factors and gas pricing behaviors. We use these features to build interpretable models using a generalizable and extensible model construction approach, in order to then discover what features best characterize transaction processing times in Ethereum, and to what extent. In particular, we discover that metrics regarding the gas price information of transactions processed in the recent past to hold the most explanatory power of all features. In comparison, the studied features which capture gas pricing behaviors hold more explanatory power than any feature dimension found in internal factors alone and combined. As a result, \dapp developers should focus on monitoring recent gas pricing behaviors when choosing gas prices for the transactions in their applications, which will ultimately help in achieving a desired level of QoS. \dapp developers should also avoid setting high gas prices that deviate too much from the recent past to avoid diminishing returns. Finally, as gas price is much more important than gas limit, \dapp developers should avoid designing contract functions which consume large amounts of gas by dividing such functions into multiple ones which consume lower amounts.

Our most robust models achieve an adjusted $R^2$ of 0.53, meaning almost half the variance is not covered by the many features we collect and engineer. As a result, we invite future work to investigate explainable models for transaction processing times which include additional features and refine the ones we propose. 

    
    
    
     

We encourage future research to use the results of our experiments to motivate the continuation of empirical studies within the area of blockchain. In relation to our study specifically, possible topics for future research include: i) the construction and interpretation of other types of models which can more accurately predict processing times, ii) similar experiments conducted in blockchain environments other than Ethereum, and iii) a similar or new set of experiments conducted for comparison which use alternative parameters to those in our study, such as time frames, statistical methods and techniques, and additional features.

	\section*{Disclaimer}
	Any opinions, findings, and conclusions, or recommendations expressed in this material are those of the author(s) and do not reflect the views of Huawei.
	
	\section*{Conflict of Interest}
	The authors declared that they have no conflict of interest. 
	
	\begin{footnotesize}
		\bibliographystyle{spbasic}      
		\bibliography{references}   
	\end{footnotesize}	

	\clearpage
	
	\appendix

	\begin{Large}
		\noindent \textbf{Appendix}
	\end{Large}
	
	\normalsize
	\vspace{-1ex}
	\section{Background}
\label{appendix:background}

In this section, we define the key concepts that are employed throughout our study.

\smallskip \noindent \textbf{Blockchain.} A blockchain is commonly conceptualized as a ledger which records all transaction-related activity that occurs between peers on the network. This activity data is stored through the usage of blocks, which is a data structure that holds a set of transactions. Each of these transactions can be identified by an identifier called the transaciton hash. Blocks are appended to one another to form a linked list type structure. A block stored in the blockchain becomes immutable once a specific number of blocks has been appended to that block. This is because for one to change the information within a specific block, they must also change all blocks appended after it as well. 

\smallskip \noindent \textbf{Transactions.} Transactions allow peers to interact with each other in the context of a blockchain. There are two types of transactions used in the Ethereum blockchain: \textit{user transactions} and \textit{contract transactions}. User transactions are employed to transfer cryptocurrency (known as Ether in Ethereum) to another user, who is identified by a unique address. Contract transactions allow users (e.g., \dapp developers) to execute functions defined in smart contracts, which are immutable, general-purpose programs deployed onto the Ethereum blockchain.

\smallskip \noindent \textbf{Gas and transaction fees.} Transactions sent within the Ethereum blockchain require a transaction fee to be paid before being sent. This fee is calculated by \textit{gas usage} $\times$ \textit{gas price}. The \textit{gas usage} term refers to the amount of computing power consumed to process a transaction. This value is constant for user transactions, costing 21 GWEI, equivalent to 2.1E-8 ETH. For contract transactions, this value is calculated depending on the bytecode executed within a smart contract, as each instruction has a specific amount of gas units required for it to execute \citep{wood2019ethereum}. The \textit{gas price} term defines the amount of Ether the sender is prepared to pay per unit of gas consumed to process their transaction. This parameter is set by the sender, and allows users to incentivize the processing of their transaction at a higher priority \citep{signer2018gas}, as miners which process the transaction are rewarded based on the transaction fee value. 

\begin{figure}[!htbp]
    \centering
    \includegraphics[width=.5\textwidth,keepaspectratio=true]{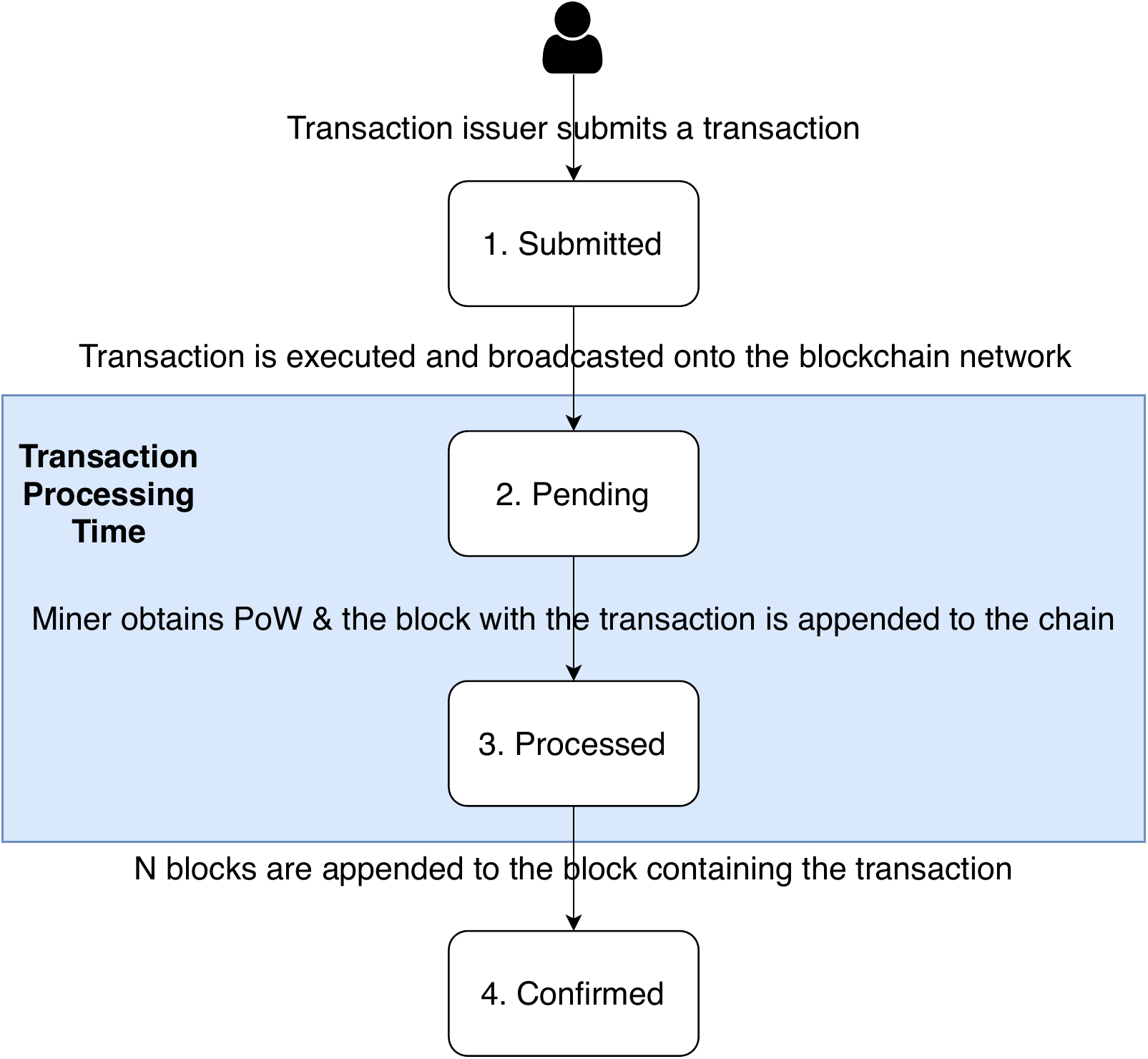}
    \caption{Transaction lifecycle in the Ethereum blockchain.}
    \label{fig:txn_lifecycle}
\end{figure}

\smallskip \noindent \textbf{Transaction processing lifecycle.} The lifecycle of a transaction transitions through various states, which are illustrated in Figure \ref{fig:txn_lifecycle}. The first state of the transaction processing lifecycle is when a user first submits their transaction $t$ to the Ethereum blockchain (\textit{submitted} state). Next, this transaction $t$ is broadcasted through the network by a peer-to-peer based communication method. Once the first node discovers transaction $t$, the transaction enters the \textit{pending} state as it enters the pending pool of transactions. As time passes this transaction $t$ is continuously discovered by an increasing amount of nodes deployed on the network, which track and record all transaction related activity, including those that are pending. Mining nodes will compete to win the Proof-of-Work (PoW)~\citep{Jakobsson99}, which will allow the winner to append their new block $b$ to the blockchain. If transaction $t$ has been processed and recorded in block $b$, transaction $t$ then transitions into the \textit{processed} state (i.e., at the \textit{block timestamp} of $b$). Transaction $t$ will transition into the next state, \textit{confirmed}, once a specific number of blocks $n$ has been appended after $b$. This number of blocks $n$ has not been officially defined, and many different values exist in various sources. For instance, $n = 7$ is stated in the Ethereum whitepaper \citep{buterin2014ethereum}. Considering that blocks are appended every 15s on average, $n = 7$ translates to approximately 1m 45s. Alternatively, some cryptocurrency exchanges define $n$ as a higher value - including Coin central, where $n$ is defined as $n = 250$, translating to approximately 1h 2m 30s \citep{coincentral}.

\smallskip \noindent \textbf{Proof-of-work.} Proof-of-Work (PoW) is a consensus protocol that requires nodes to solve a difficult mathematical puzzle. This protocol ensures that the strategy used to solve the puzzle is optimal (i.e. no solution for the puzzle exists other than a brute force solution). Fortunately the verification of such a solution is much cheaper in terms of computation. Ultimately this allows the nodes, or entities within the network who do not know the authority of one another, to trust each other and build a valid transaction ledger without requiring a centralized third-party authority (e.g., a bank).

\smallskip \noindent \textbf{Next-generation \dapps.} Commonly, \dapps expose the internal blockchain complexities to end-users, ultimately requiring that they install a \textit{wallet} (e.g., Metamask). A wallet is a computer program (commonly a web browser extension) that enables one to submit transactions in a blockchain without having to set up their own node. However, submitting transactions with a wallet still requires end-users to set up \textit{transaction parameters}. Such a task is error-prone by construction~\citep{Oliva21}, as not all end-users are familiar with blockchain concepts and their intricacies. Other problems include end-users getting confused with the user interface of wallets or even making typos while inputting the transaction parameter values. As an illustrative example, an Ethereum user called Proudbitcoiner spent 9,500 USD in transaction fees to send just 120 USD to a smart contract\footnote{The transaction logs can be seen at \url{https://etherscan.io/tx/0x5641c38bf30fd19ad3332100762017573ee676f35250dcf4a976bb0cfe31ac2f}}. The user said the following (\textit{gas limit} and \textit{gas price} are two transaction parameters and 1 GWEI = 1E-9 ETH): \label{appendix:subsec:nextgendapps}

\begin{quote}
    \small \textit{``Metamask didn't populate the 'gas limit' field with the correct amount in my previous transaction and that transaction failed, so I decided to change it manually in the next transaction (this one), but instead of typing 200000 in ``gas limit'' input field, I wrote it on the ``Gas Price'' input field, so I paid 200000 Gwei for this transaction and destroyed my life.''}
\end{quote}

Due to incidents such as the one above, alongside end-user onboarding barriers, more recently \dapp developers have started to engineer their applications using a new paradigm, which hides all blockchain complexities from end-users. In this new paradigm, the burden of setting up transaction parameters is on the \dapp developers. \dapp developers, in turn, can charge end-users with a flat fee (e.g., payable via credit card), such that they do not need to hold Ether or use wallets, ultimately easing their onboarding process. We refer to these \dapps as the next-generation \dapps. Next-generation \dapps are a reality and  include \dapps such as MyEtherWallet\footnote{\url{https://www.myetherwallet.com/}}.

\clearpage

\section{Model Construction: Correlation and Redundancy Analysis}
\label{appendix:corr_redun_analysis}

\begin{table}[H]
    \centering
    \caption{The independent features removed resulting from conducting the correlation analysis in Section~\ref{sec:model-construction}}
    \label{tab:corr_analysis}
    \begin{tabular}{lll} \hline
    \textbf{Index} &    \textbf{Feature 1 (Removed)}                    & \textbf{Feature 2}           \\ \hline
    1&    num\_same\_120                                    & avg\_num\_same\_120 \\
    2&    difficulty\_avg\_120                              & difficulty\_med\_120       \\
    3&    med\_num\_above\_120                              & num\_above\_120\\
    4&    num\_below\_120                                   & avg\_num\_below\_120 \\
    5&    avg\_num\_above\_120                              & med\_num\_above\_120 \\
    6&    avg\_num\_below\_120                              & med\_num\_below\_120\\
    7&    total\_txs\_120                                   & avg\_txs\_120\\
    8&    difficulty\_1                                     & difficulty\_med\_120\\
    9&    avg\_func\_gas\_usage\_120                        & med\_func\_gas\_usage\_120\\
    10&   avg\_txs\_120                                     & med\_txs\_120\\
    11&   avg\_num\_same\_120                               & std\_num\_same\_120\\
    12&   avg\_pending\_pool\_120                           & med\_pending\_pool\_120\\
    13&   avg\_gas\_price\_1                                & med\_gas\_price\_1\\
    14&   avg\_gas\_price\_120                              & med\_gas\_price\_120\\
    15&   med\_num\_above\_120                              & med\_num\_below\_120\\
    16&   med\_gas\_price\_120                              & med\_gas\_price\_1\\
    17&   pending\_pool                                     & med\_pending\_pool\_120\\
    18&   med\_gas\_price\_120                              & gas\_price\_gwei \\ 
    19&   input\_length                                     & to\_contract\\
    20&   to\_contract                                      & contract\_bytecode\_length \\
    21&   avg\_gas\_price\_gwei\_prev\_day                  & med\_gas\_price\_120 \\
    22&   past\_avg\_mins\_120                              & past\_std\_mins\_120\\
    23&   med\_func\_gas\_usage\_120                        & std\_func\_gas\_usage\_120 \\
    24&   contract\_bytecode\_length                        & std\_func\_gas\_usage\_120\\
    25&   num\_same\_1                                      & pct\_same\_1 \\
    26&   pct\_above\_120                                   & med\_pct\_above\_120\\
    27&   avg\_pct\_above\_120                              & med\_pct\_above\_120\\
    28&   pct\_below\_120                                   & avg\_pct\_below\_120\\
    29&   avg\_pct\_below\_120                              & med\_pct\_below\_120\\
    30&   med\_num\_same\_120                               & med\_pct\_same\_120\\
    31&   pct\_same\_120                                    & avg\_pct\_same\_120\\
    32&   med\_pct\_above\_120                              & med\_pct\_below\_120 \\
    33&   std\_num\_same\_120                               & std\_pct\_same\_120 \\
    34&   med\_num\_below\_120                              & med\_pct\_below\_120\\
    35&   std\_pct\_above\_120                              & std\_pct\_below\_120  \\
    36&   avg\_pct\_same\_120                               & std\_pct\_same\_120\\
    37&   num\_above\_1                                     & pct\_above\_1     \\
    38&   num\_below\_1                                     & pct\_below\_1     \\
    39&   std\_num\_above\_120                              & std\_pct\_below\_120 \\
    40&   pct\_above\_1                                     & pct\_below\_1 \\
    41&   pct\_below\_1                                     & med\_pct\_below\_120\\ 
    42&   num\_pending                                      & med\_pend\_prices \\ 
    43&   avg\_pend\_prices                                 & med\_pend\_prices \\
    44&   std\_pend\_prices                                 & med\_pend\_prices\\
    \hline
    \end{tabular}
\end{table}


\clearpage

\section{SHAP}
\label{appendix:shap}

\begin{figure}[H]
    \centering
    \includegraphics[width=\linewidth]{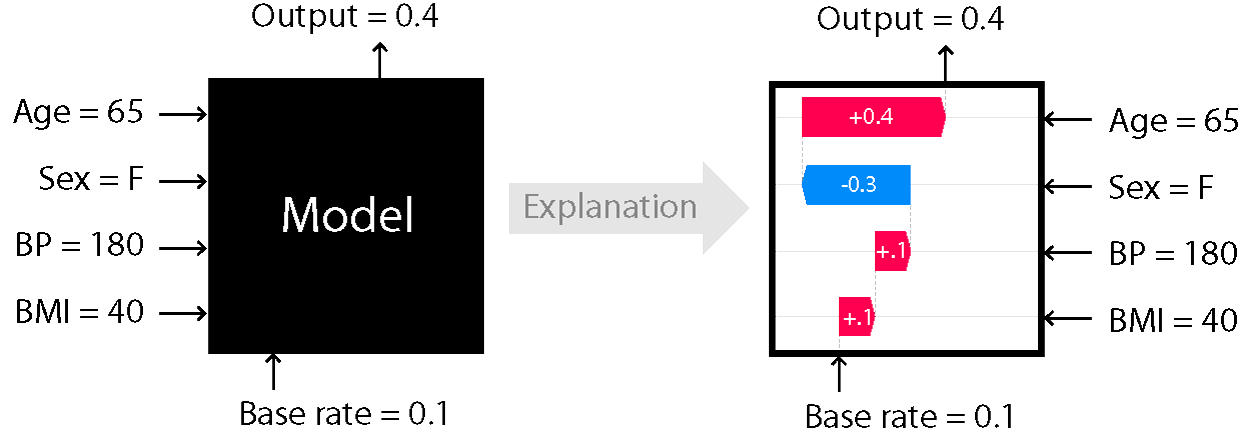}
    \caption{Intuition behind the SHAP model interpretation technique (adapted from \url{https://shap.readthedocs.io}).}    
    \label{fig:shap-idea}
\end{figure}

To interpret our random forest models at the feature level, we use a model-agnostic interpretation technique called SHAP~\citep{Lundberg17}. The intuition behind SHAP is shown in Figure~\ref{fig:shap-idea}. Assume that we are using a black-box model to predict the chances of someone eventually having a heart attack (HT). The model's features are Age, Sex, BP (blood pressure), and BMI (body mass index). By taking the mean of the HT column, we have an expected value for HT. We refer to this expected value as the \textit{base rate}. In the example shown in Figure~\ref{fig:shap-idea} (left-hand side), this base rate is 0.1 (10\%). For a particular instance (row) in the dataset (Age = 75, Sex = F, BP = 180, and BMI = 40), the black-box model outputs a HT of 0.4. The question now is: how much does each feature contribute to moving the output from 0.1 (base rate) to 0.4 (the actual prediction)? In an additive model such as a linear model, the effect of each feature is simply the weight of the feature times the feature value. SHAP relies on game theory to extract the features contributions from any machine learning model. As shown in the right-hand side of Figure~\ref{fig:shap-idea}, SHAP indicates that BMI = 180 contributed with +0.1 to the base rate (positive impact, shown in red), BP = 180 contributed with an additional +0.1 to the base rate, Sex = F contributed with -0.3 to the base rate (negative impact, shown in blue), and Age = 65 contributed the most with +0.4 to the base rate. If we sum up all these individual feature contributions, we obtain the final prediction of HT = 0.4. Hence, SHAP is additive in nature and can easily explain how each feature contributed to the prediction of a given instance in the dataset. SHAP has been increasingly adopted in the Software Engineering community~\citep{Viggiato20, Esteves20, Gopi21}. 

\end{document}